\documentclass[3p,authoryear]{elsarticle}
\usepackage[dvipsnames]{xcolor}
\usepackage{nohyperref}
\usepackage{amsmath,amscd,amssymb,graphicx,overpic,multicol,multirow,booktabs,natbib,setspace,tikz,cleveref,array}
\usepackage[T1]{fontenc}
\usepackage{float, caption, subcaption} 
\usepackage{url}

\usetikzlibrary{backgrounds,shapes,arrows,positioning,calc}
\graphicspath{{figures/}}
\crefname{section}{\S}{\S}

\usepackage{palatino}
\usepackage{textcomp}
\definecolor{MatlabCellColour}{RGB}{250,250,250}
\definecolor{MatPurp}{rgb}{.625,.1406,.9375}
\usepackage{listings}

\lstdefinestyle{customc}{
	belowcaptionskip=.25\baselineskip,
	breaklines=true,
	frame=L,
	xleftmargin=\parindent,
	language=Matlab,
	showstringspaces=false,
	basicstyle=\small\ttfamily,
	keywordstyle=\bfseries\color{white!30!black},
	identifierstyle=\color{blue},  
	commentstyle=\itshape\color{green!60!black},
	stringstyle=\color{MatPurp},
	backgroundcolor=\color{MatlabCellColour}
}

\newcolumntype{L}[1]{>{\raggedright\let\newline\\\arraybackslash\hspace{0pt}}m{#1}}
\newcolumntype{C}[1]{>{\centering\let\newline\\\arraybackslash\hspace{0pt}}m{#1}}
\newcolumntype{R}[1]{>{\raggedleft\let\newline\\\arraybackslash\hspace{0pt}}m{#1}}

\newcommand{\td}[1]{\tilde{#1}}

\DeclareMathOperator*{\argmin}{arg\,min}

\newcommand{\bPsi}{\boldsymbol{\Psi}}
\newcommand{\bPhi}{\boldsymbol{\Phi}}
\newcommand{\bTheta}{\boldsymbol{\Theta}}
\newcommand{\bphi}{\boldsymbol{\phi}}
\newcommand{\bSigma}{\boldsymbol{\Sigma}}
\newcommand{\bS}{\mathbf{S}}
\newcommand{\bP}{\mathbf{P}}
\newcommand{\bs}{\mathbf{s}}

\newcommand{\bV}{\boldsymbol{V}}
\newcommand{\ff}{\mathbf{f}}

\newcommand{\reals}{\mathbb{R}}
\newcommand{\ba}{\mathbf{a}}
\newcommand{\bX}{\boldsymbol{X}}

\newcommand{\ie}{{\em i.e.}}
\newcommand{\bt}{\mathbf{t}}
\newcommand{\bx}{\mathbf{x}}
\newcommand{\by}{\mathbf{y}}
\newcommand{\bz}{\mathbf{z}}

\newcommand{\bQ}{\mathbf{Q}}
\newcommand{\bR}{\mathbf{R}}

\usepackage{etoolbox}
\newcommand\revtwo[3][]{%
	\ifstrempty{#1}{}
	{}%
	\ifstrempty{#3}{{\color{black}{#2}}}%
	{{\color{black}{#2}}}
}

\newcommand\revision[3][]{%
\ifstrempty{#1}{}
{\textcolor{black}{}}
\ifstrempty{#3}{{\color{black}{#2}}}%
{{\color{black}{#2}}}
}

\tikzstyle{block} = [draw, fill=gray!20, rectangle, 
    minimum height=3em, minimum width=3em,rounded corners]
\tikzstyle{sum} = [draw, fill=blue!20, circle, node distance=1cm]
\tikzstyle{input} = [coordinate]
\tikzstyle{output} = [coordinate]
\tikzstyle{pinstyle} = [pin edge={to-,thin,black}]
\tikzstyle{pinstyle2} = [pin edge={to-,thin,black}]
\tikzstyle{tmp} = [coordinate]

\title{\LARGE{\textbf{Environment Identification in Flight using Sparse Approximation of Wing Strain}}}

\author[uwamath]{Krithika Manohar\corref{cor}} \ead{kmanohar@uw.edu}
  \cortext[cor]{Corresponding author. \emph{E-mail:} kmanohar@uw.edu} 
  \author[uwme]{Steven L. Brunton}
  \author[uwamath]{J. Nathan Kutz} 
  \address[uwamath]{Department of
    Applied Mathematics, University of Washington, Seattle, WA 98195,
    United States}
    \address[uwme]{Department of
    Mechanical Engineering, University of Washington, Seattle, WA 98195,
    United States}
    
\makeatletter
\def\blfootnote{\xdef\@thefnmark{}\@footnotetext}
\makeatother

\begin{document}
{*Accepted for publication in Journal of Fluids and Structures, January 2017 \\ \verb+10.1016/j.jfluidstructs.2017.01.008+}

\blfootnote{\textcopyright~2016. This manuscript version is made available under the CC-BY-NC-ND 4.0 license \url{http://creativecommons.org/licenses/by-nc-nd/4.0/}}
 \begin{abstract}

 This paper addresses the problem of identifying different flow environments from sparse data collected by wing strain sensors. Insects regularly perform this feat using a sparse ensemble of noisy strain sensors on their wing.
 First, we obtain strain data from numerical simulation of a {\em Manduca sexta} hawkmoth wing undergoing different flow environments.
Our data-driven method learns low-dimensional strain features originating from different aerodynamic environments using proper orthogonal decomposition (POD) modes in the frequency domain, and leverages sparse approximation  to classify a set of strain frequency signatures using a dictionary of POD modes. This bio-inspired machine learning architecture for dictionary learning and sparse classification permits fewer costly physical strain sensors while being simultaneously robust to sensor noise. A \revtwo{measurement selection}{sensor placement} algorithm identifies frequencies that best discriminate the different aerodynamic environments in low-rank POD feature space. In this manner, sparse and noisy wing strain data can be exploited to robustly identify different aerodynamic environments encountered in flight, providing insight into the stereotyped placement of neurons that act as strain sensors on a {\em Manduca sexta} hawkmoth wing.

\end{abstract}

\begin{keyword}
insect flight; \revision[]{proper orthogonal decomposition}{dictionary learning}; sparse approximation; \revision{classification}{compressed sensing}; sensor selection; unsteady aerodynamics 
\end{keyword}

\maketitle
\setlength{\parskip}{5pt}
\section{Introduction}

Winged flight remains one of the most successful forms of animal locomotion and one of mankind's foremost accomplishments. The aircraft of today are a direct result of centuries of fascination, inquiry and experimentation inspired by bat, bird and insect locomotion. Today's most prevalent rigid-wing aircraft perform maneuvers and functions strikingly different from that of these flexible-winged animals, and the success of these aircraft in military and transport continues to spur development of rigid-wing technology. More recently, advances in robotics, materials and high-performance computation have advanced bio-inspired flexible-wing technologies including miniaturized unmanned micro-aerial vehicles (MAVs), ornithopters, hovercraft, and drones. Suitable controllers for these small-scale autonomous technologies require extensive understanding of low-Reynolds number, unsteady aerodynamics that is often experienced by bats, insects and most birds.  Of particular 
interest in this work is understanding the role that a limited number of sensors (e.g. neurons) play in accurately informing control decisions in this low-Reynolds number regime, thus potentially helping
to reveal bio-inspired flight control principles.

While the comprehensive mechanism of wing actuation, fluid-structure coupling and response maneuvers in winged animal flight is not fully understood,  biologists and engineers have nevertheless made significant progress in identifying the propulsive forces in flapping-wing flight \citep{Ellington:1999,Zbikowski:2002,Tangorra:2007,Dabiri:2009}. The preliminary study of these forces began with quasi-steady, inviscid assumptions from rigid-wing thin airfoil theory that proved to be inadequate at predicting the additional lift generated by insects in experiments. It was later found that the unsteady aerodynamics and added-mass of the surrounding fluid are crucial for characterizing the forces experienced by animal wings \citep{Ellington:1994,Sane:2003,Wang:2005}. Although the fluid's added-mass would seem to complicate the robust generation of lift, it was discovered that animal wings rotate and flap in a manner that harness aerodynamic added-masses and leading edge vortices for additional lift in insects \citep{DickinsonSane:1999,Birch:2001,Combes:2001,SongBreuer:2008,Eldredge:2010,faruque2010insect,faruque2010insect_2}, birds \citep{Spedding:2003}, and bats \citep{Hedenstrom:2007,ClarkSmits:2006}. \revision[R2:1]{There is evidence that wings harness low-dimensional flow structure such as leading edge vortices to maximize lift and improve stability in flight \citep{Birch:2001,videler2004leading,Dabiri:2009}. This is consistent with the observation that wing motion is itself low-dimensional.}{It is probable that wing motions harness a low-dimensional flow structure, as the wing motion is itself low-dimensional:} Proper orthogonal decomposition analysis has revealed that fin motion \citep{Tangorra:2007}, bat \citep{Riskin:2008} and avian wing motion are constrained to only a few degrees of freedom.  The lack of complex musculature in insect wings results in even fewer degrees of freedom.  Such dimensionality reduction suggests that insects, which are constrained to small ranges of motion, capitalize on low-dimensional feature spaces \revision{to inform}{for informing} low-dimensional control protocols.

Wing mechanosensors in insects detect and leverage these \revision[R2:2]{fluid forces and accelerations}{large fluid features} for changing flight environments. Indeed, tactile sensory mechanisms have been identified on the wings of most animals - bats sense with small hairs on their wings \citep{Sterbing:2011}, birds sense with wing feathers \citep{Brown:1993}, but insects possess a small number of strain sensors on their wings called {\em campaniform sensilla}. The {\em sensilla} are strongly implicated in neurosensory flight control \citep{Dickerson:2014,Sane:2007}, in part because insect wings react to disturbances faster than visual stimulus transmission to the central nervous system \citep{Collett:1975}. \revision[R2:6]{Biological evidence shows that sensilla are stereotyped across specimens of the same species. Indeed, \cite{cole1982pattern} show that the spatial distribution of sensilla are encoded in the genes of the fruit fly {\em Drosophila}, and sensilla determine maneuvers in flight control in locusts and {\em Manduca sexta} \citep{gettrup1966sensory,Dickerson:2014,dickerson2015role}.}{} Aerodynamic feedback is encoded within the strain signals, registering fluidic loading frequencies into the signal that can be exploited for characterizing flow environment. \revision[R2:3]{Evidence shows}{it is probable} that insects exploit innate or learned knowledge of fluid environment through the strain encodings to make split-second decisions in flight. The insect nervous system has evolved specifically for the decision task, but controllers in hovercraft and MAVs require other means of exploiting point sensor feedback. Strain sensors are too sparsely distributed to fully resolve spatial flow encoding over the wing, and equation-based flow identification or prediction is expensive and difficult to generalize to different flow regimes. 

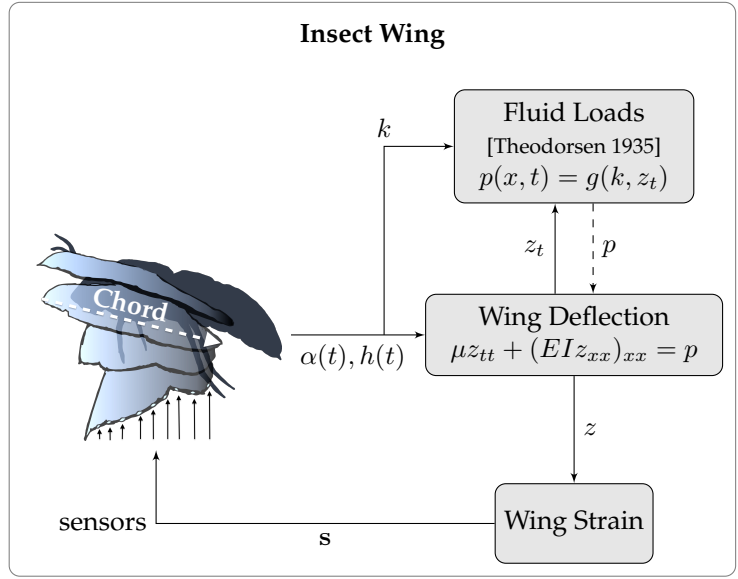
\begin{figure}
	\centering
	\begin{tikzpicture}[framed,background rectangle/.style={draw=gray,rounded corners},auto, node distance=2cm,>=latex']
	\node [name=input] 
		{\begin{overpic}[width=.2\textwidth]
	    {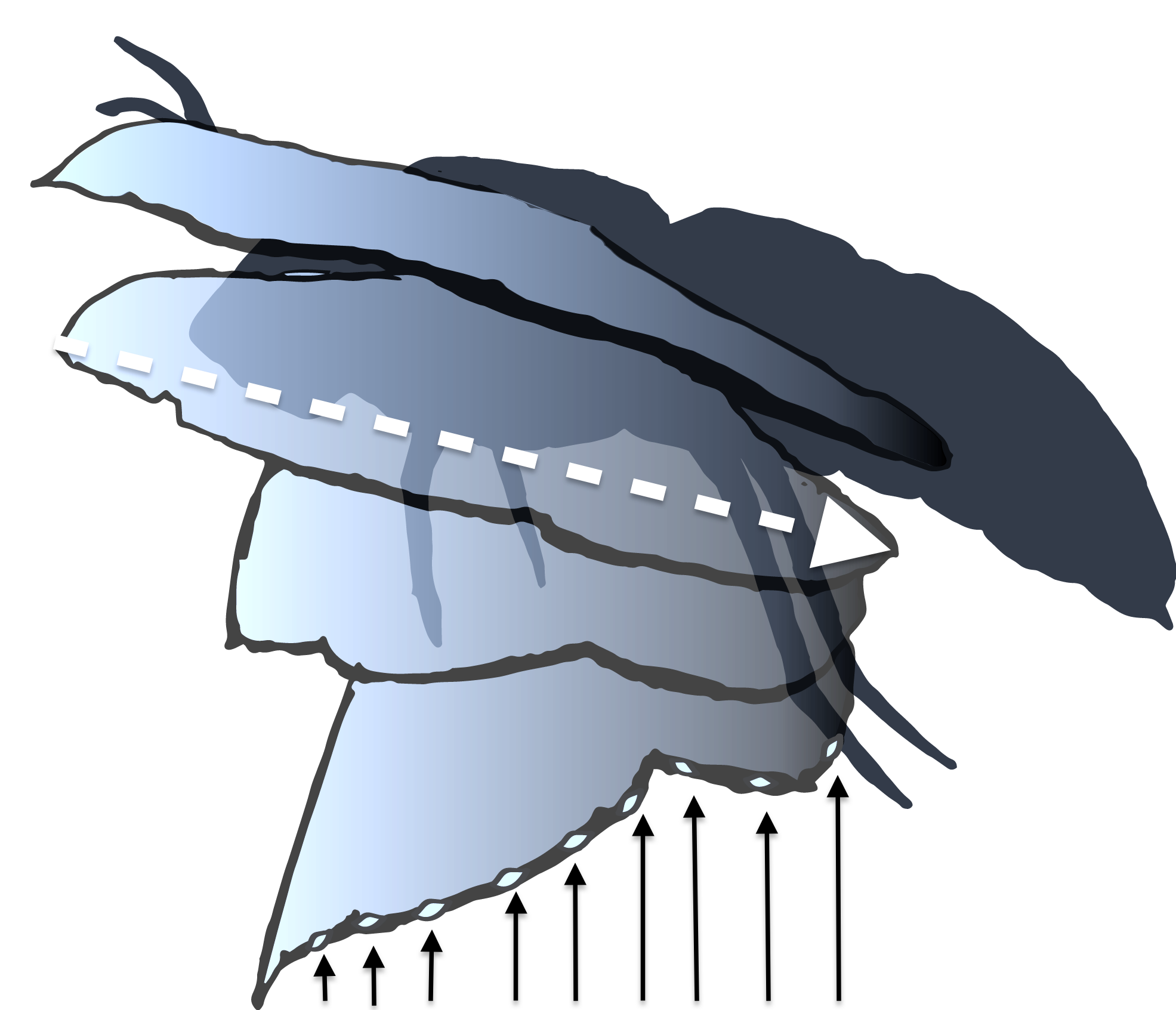}
	    \put(40,55){\makebox(0,0){\rotatebox{-14}{\color{white}\bf Chord}}}
	    \end{overpic}};
	\node [block, right of=input, node distance=5.5cm] (beam) 
	    {\begin{tabular}{c} Wing Deflection \\  $\mu z_{tt}+(EI z_{xx})_{xx}= p$\end{tabular}};
	\node [block, above of=beam,node distance=2.5cm] (fluid) 
	    {\begin{tabular}{c} Fluid Loads \\ {\footnotesize [Theodorsen 1935]} \\$p(x,t)=g(k,z_t)$\end{tabular}};
	\node [block, below of=beam,node distance=2.5cm] (strain) {Wing Strain};
	\node [tmp, below of=input,node distance=2.5cm] (tmp1){$H(s)$};
	\node [tmp, left of=beam,node distance=2.5cm] (tmp2){$H(s)$};
	\draw [draw,->] (input) -- node[anchor=north,xshift=.2cm] {$\alpha(t),h(t)$} (tmp2) -- (beam) ;
	\draw [draw,->] (tmp2)  |-  node{$k$} (fluid) ;
	\draw [draw,->] ($(beam.north)-(0.25,0)$)-- node {$z_t$} ($(fluid.south)-(0.25,0)$);
	\draw [dashed,->] ($(fluid.south)+(0.25,0)$)-- node {$p$} ($(beam.north)+(0.25,0)$);
	\draw [draw,->] (beam) -- node {$z$} (strain);
	\draw [draw,->] (strain)-- node{$\bs$}(tmp1) -| node{sensors}(input);
	\node[align=center,font=\bfseries, yshift=2em] (title) 
	    at (current bounding box.north)
	         {Insect Wing};
	\end{tikzpicture}
	\caption{Strain sensors in insect wings directly encode elastic wing deformation from the leading edge dynamics and flow feedback loads. These fluid loads are driven by dominant flapping and external reduced frequencies $k$ and weak coupling from wing deformation. \label{fig:airfoilwingcomp}} 
\end{figure}
 
\revision[R2:4]{In contrast, data-enabled methods for flow characterization or parameter estimation have shown remarkable promise in the analysis of complex flows. A variety of data decompositions are being applied to flow measurement data for spectral analysis, model reduction, and control of complex flows, for example, proper orthogonal decomposition (POD)~\citep{lumley1970,holmes1998}, dynamic mode decomposition (DMD)~\citep{dmd1,dmd2,dmd3,dmd4,Kutz2016book}, compressed sensing \citep{Bright:2013,bai2014low,Brunton:2014} and sparse regression~\citep{Brunton2016pnas}, network theoretic approaches \citep{nair2015network,Taira2016jfm}, as well as many other machine learning methods for flow control, as surveyed in \cite{brunton2015survey}. Broadly, these equation-free methods characterize measurements of a system to inform its state and subsequent control decisions. In addition to intriguing evidence that biological strain sensors inform reactive decisions in flight, sensors play a pivotal role in the feedback control of many complex flows. }
{Data-assisted approaches are more flexible for learning and differentiating flow features for efficient flow identification.} 

\subsection{Contributions of this work}
The focus of our work is sensor-enabled, data-driven flow environment classification using supervised learning and sparse approximation of incoming sensor measurements. 
Our data consists of strain point measurements from a numerical fluid-structure interaction model of a hawkmoth wing undergoing different environments of flow feedback. 
First, POD modes of strain measurements from different aerodynamic environments are assembled into a dictionary of low-rank dynamical states labelled by environment. This supervised learning stage mimics experiential learning and trains the data-driven model specifically for the task at hand. Next, incoming strain measurements are classified in this learned POD library using $L_1$ constrained sparse approximation and sparse representation for classification (SRC)~\citep{Wright:2009}. 
\revision[R1:8]{We propose a framework in which sparse approximation is used to classify flow environments. Although other classifiers such as discriminant-based analysis and neural networks abound in machine learning, sparse approximation is particularly suited for the problem at hand since it  is naturally robust to noise in measurement and accommodates classification between subspaces of different dimension. Moreover, sparse approximation aligns with plausible neurobiological sparse encoding strategies \citep{Olshausen:2004}.}{}

In this work, sparse approximation of wing strain for flow classification is shown to be accurate and robust to high levels of sensor noise. \revision[R2:5]{Our strain measurements consist of Fourier coefficients in the frequency domain at a single spatial location on the wing chord. These frequency measurements are then sparsely approximated in an overcomplete library of POD bases that characterize different aerodynamic environments. An $L_1$ sparsity constraint is imposed on the solution so that the nonzero solution components identify the originating flow environment.}{Our measurement data are the strain frequency content at one location on the wing, and the dictionary of reduced POD bases accordingly span this frequency content.} This analysis of strain time histories in the frequency domain permits the use of limited spatial sensors and \revision{provides robustness}{is robust} to sensor noise. \revtwo[R2:2]{Frequencies are sampled from a single, randomly chosen spatial location at a time. We also perform a location study in which the classification accuracy is compared across different regions along the chord. Based on this comparison, trailing edge frequency content is amplified and more suitable for aerodynamic environment discrimination. This is because the more pliable trailing edge contains more energetic aerodynamic contributions far away from leading edge actuation dynamics.}{} Sparse representation accuracy suggests sampling certain frequencies yields higher flow classification accuracy, and \revtwo[R2:1]{measurement}{sensor} selection algorithms are used to \revision{identify these discriminating frequencies}{determine these selected frequencies}.  This further suggests that a principled \revtwo{measurement strategy}{approach to sensor placement} on flexible wings is advantageous, and is consistent with the stereotyped placement of strain-sensing neurons in the hawkmoth.



\begin{figure*}[t]
	\setlength{\fboxrule}{0.5pt}%
\framebox{
\noindent\begin{minipage}[h]{.48\columnwidth}
\section*{Nomenclature}
\begin{tabular}{ll}
$\ba$ & Vector of POD coefficients \\
$b$ & Half-chord length of wing [.01 m] \\
$c$  & Categories or number of environments \\
$C(k)$ & Theodorsen transfer function \\
$EI(x)$ & Flexural stiffness of wing [Nm$^2$]\\ 
$f$ & Frequency of maneuver [Hz]\\
$\ff$ & Vector of frequencies \\
$k$ & Reduced frequency [$k\triangleq{2\pi f b}/{U_{\infty}}$]\\
$h_0$ & Initial vertical position of wing\\
$h(t)$ & Vertical position of plate\\
$m$ & Spatial grid resolution \\
$N$ & Number of timesteps \\
$p(x,t)$ & Chordwise loading of fluid pressure \\
$\bP$ & Measurement selection matrix \\
$r$ & Reduced-order model order \\
$\bs$ & Vector of transverse strain of wing \\
$\bS$ & Matrix of transverse strains \\
$\hat\bS$ & Discrete Cosine Transform of $\bS$ \\
$t$ & Time [s] \\
\end{tabular}
\end{minipage}
\begin{minipage}[h]{.48\columnwidth}
\begin{tabular}{ll}
$\mathbf{t}$ & Vector of time $t$ in seconds \\
$U_\infty$ & Free stream velocity [10 m/s]\\
$\mathbf{V}$ & Matrix of right singular vectors of $\bS$ \\
$x$ & Chordwise spatial coordinate [m] \\
$\bx$ & \revision{High-dimensional signal}{Vector of spatial coordinates $x$} \\
$z(x,t)$ & Transverse deflection of wing [m] \\
$\alpha_0$ & Base angle of attack\\
$\alpha(t)$ & Angle of attack of wing \\
$\omega$ & Angular velocity of maneuver \\
$\mu$ & Linear wing density [.002 kg/m]\\
$\xi$ & Zero-mean sensor noise\\
$\eta$ & Sensor noise variance \\
$\epsilon$ & Error tolerance parameter \\
$\sigma$ & Singular values of POD decomposition \\
$\bSigma$ & Diagonal Matrix of singular values \\
$\bPhi$ & Basis of POD modes \\
$\bPhi_r$ & Low-rank basis of $r$ dominant POD modes  \\
$\hat\bPhi_r$ & $r$ dominant POD modes of $\hat\bS$  \\
$\bPsi,\hat\bPsi$ & Library of multiple low-rank POD bases \\
$\bTheta_r$ & Joint POD modes of all available data
\end{tabular}
\end{minipage}}
\end{figure*}

\section{Background}

This section is a brief overview of existing literature on lift generation in insect flight, the role of insect wing mechanosensors, and data-driven sparse approximation methods. \revision{We also introduce important aerodynamic flow and actuation parameters that will be used throughout the paper.}{}

\subsection{Unsteady aerodynamics of insect flight}
The uniqueness of insect flight can be attributed to simple oscillating wing actuation (ranging from 10 to 500 Hz) that can achieve a large variety of maneuvers in exceedingly short timescales. Indeed, a wider range of motion is accessible to insects from relatively simpler wing actuation compared to birds or fixed wing structures. Unlike birds, insects also possess little to no wing musculature to control wing shape and do not require the large forward velocities of the fixed wings in airplanes. The study of insect wings is extremely relevant to miniaturized flight applications for unmanned MAVs that are required to perform sophisticated maneuvers within small spatial and temporal scales. The rapid miniaturization of these technologies in surveillance and robotics has in part fueled the study of lift generation in insects.

\revision[R1:1,R2:9]{Modeling the forces and moments acting on an insect wing is challenging because the rapid small scale movement is highly {\em unsteady}, meaning wings maneuver rapidly enough that excited vortices do not have time to convect the entire chord length before affecting the wing. In contrast, steady or quasi-steady aerodynamics typical of rigid airfoils rest on the assumption that maneuvers occur slowly relative to the fluid convection time.  
A comprehensive treatment of unsteady aerodynamics can be found in \cite{Leishman:2006}, including the extensively used classical models of \cite{wagner1925} and \cite{Theodorsen:1935} (which we subsequently use in modeling fluid-structure interaction). Unsteady flight is commonly parametrized by the Strouhal number $St=fM/U_\infty$ and reduced frequency of oscillation, $k=2\pi f b/U_\infty$, where $f$ and $M$ are the frequency and amplitude of oscillation, $b$ is the half-chord length, and $U_\infty$ is the free-stream velocity. In particular, gusts, flow disturbances and rapid maneuvers in unsteady regimes excite higher frequencies resulting in large reduced frequencies $k$ greater than 1.
The fluid-structure interaction in our work models a wide range of reduced frequency regimes to be subsequently characterized by sparse classification based on limited frequency measurements.}

The unsteady effects, forces and lift generation experienced by insect wings have been extensively studied by \cite{Ellington:1994}, \cite{Wang:2005} and \cite{DickinsonSane:1999}, among others.  These seminal works in the study of insect wings all argue that unsteady effects are crucial for the additional lift generation previously unaccounted for by quasi-steady assumptions in thin airfoil theory. Although thin airfoil theory presents a starting point in characterizing surrounding flows, the fluid-structure interactions in airfoils and insect wings are markedly different. Airfoils respond to resonant frequencies between actuation and vortex shedding, and this is the focus of aeroelasticity and wing flutter analysis~\citep{dowell:1979,Dowell:1996,dowell:2001}. Insect wings, however, are strained and bent in response to both inertial-elastic forces and fluid loads which can be weakly coupled as shown in Figure \ref{fig:airfoilwingcomp}. This complex interaction between elastic bending and fluid forcing is used in our simulations of the wing and the subsequent strain data.

\subsection{Theodorsen model: unsteady forces on wings}

\cite{Theodorsen:1935} addresses airfoil aerodynamic instability and flutter on thin plates undergoing oscillatory pitching and plunging movement. The Theodorsen model yields lift forces, pressures and moments of a rigid airfoil characterized by oscillatory movement in the angle of attack $\alpha(k)$ and vertical position $h(k)$. Its predictions agree quite well with experimental results \citep{Leishman:2006,Brunton2013jfm} compared to the predictions of quasi-steady thin airfoil theory. This improvement is because Theodorsen augments the quasi-steady lift term from thin airfoil theory \revision[R2:8]{\citep{Leishman:2006}}{},
\begin{equation}
\label{eqn:quasisteady}
C_L = \frac{2\pi}{U_\infty} \left[U_\infty\alpha+\dot{h}+\dot{\alpha}\left(\frac{1}{2}-\frac{b}{2}\right)\right],
\end{equation}
\revision{with unsteady terms multiplied by the Theodorsen transfer function, $C(k)$, which is parametrized by reduced frequency of plate oscillation $k$}{which is known as the Theodorsen transfer function and depends on the reduced frequency $k$}. The transfer function $C(k)$ is derived by integrating forces generated by planar wake vorticity beyond the foil in inviscid, incompressible flow. We compare the quasi-steady lift coefficient \eqref{eqn:quasisteady} to the Theodorsen lift coefficient and distinguish added-mass and circulatory terms \revision[R2:8]{\citep{Leishman:2006}}{}:
\begin{equation}
\label{eqn:theodorsen_lift}
C_L = \frac{b\pi}{U_\infty^2} [U_\infty\dot{\alpha}  + \ddot{h}-\frac{b^2}{2}\ddot{\alpha}]-\frac{2\pi}{U_\infty} C(k)\left[U_\infty\alpha+\dot{h}+\dot{\alpha}\left(\frac{1}{2}-\frac{b}{2}\right)\right].
\end{equation}
$C(k)$ is a quotient of Hankel functions of the second kind
\begin{equation}
C(k) = \frac{H_1^{(2)}(k)}{H_1^{(2)}(k)+iH_0^{(2)}(k)},
\end{equation}
where the subscripts of the Hankel function denote its order.

The model is an invaluable tool for understanding fluid-structure coupling along insect wings because it is based on unsteady assumptions and incorporates added-mass of the fluid surrounding the wing. Although the model rests on inviscid assumptions, \cite{Brunton:2013} empirically adjusted coefficients in the model to agree with low-Reynolds number regimes typical of insect flight. The Theodorsen model will be used in our analysis for generating libraries of POD modes characterizing different aerodynamic environments parametrized by different $k$.

\subsection{Wing strain}
Strain sensors encode the unsteady fluid added-masses on the wing and capture elastic stretching due to inertial-elastic forces from the flapping movement. Surprisingly, 80\% of the elastic wing deflection in insects is due to inertial elastic forces \citep{Combes:2003,Eberle:2014}, and the direct deformations due to flow feedback are relatively small in magnitude. \revision[R2:7]{Insects nevertheless appear to characterize flow environments using limited sensory information resulting from fluid forces upon the wing. Our fluid-structure interaction model accordingly simulates the wing using Theodorsen's fluid pressures as a forcing term to a wing undergoing inertial-elastic deformation.}{Strain is a direct byproduct of deflection, and so it is surprising that insects perform robust environment identification when environmental feedback comprises less than a quarter of the available information. 
This is because wing dynamics are dominated by elasticity and bending from small magnitude forcing of flow feedback.} To obtain strain data, we numerically simulate the wing chord deformation $z(x,t)$ using classical elastic beam theory \citep{GuentherLee:1988} with external aerodynamic forcing term $p(x,t)$ from Theodorsen's chordwise loads,
\begin{align}
\label{eqn:beamgeneral}
 &\mu z_{tt} + (EI z_{xx})_{xx} = p(x,t), ~~x\in(0,2b),t\ge 0 .
\end{align}
Parameters such as the flexural stiffness $EI(x)$ in both the beam equation and the Theodorsen model reflect the wing morphological traits and leading edge flapping frequency of the representive hawkmoth insect. 
The chordwise spatial wing strain can be easily calculated from the structural deformation $z(x,t)$ as the change in chord length divided by the initial length. \revtwo[R2:3]{We emphasize, however, that strain {\em measurements} used for classification are Fourier coefficients of the chordwise strain temporal dynamics at a single spatial location. Ultimately sparse approximation classifies sampled frequencies based on existing strain knowledge forced by different $p(x,t)$. }{Ultimately we use this theoretical construct to generate strain data, take discrete strain measurements (point sensors), and classify sampled frequencies based on existing strain knowledge forced by different $p(x,t)$ using sparse approximation.}

\subsection{Exploiting sparsity for classification}

Most organisms, including insects, exploit low-dimensional structure to quickly characterize sensory input and execute swift response. They rely on innate or learned knowledge of new regimes of sensory stimuli to characterize stimuli on the fly. A data-driven protocol requires a similar means of acquiring low-rank knowledge from wing strain measurements. Many classifiers, including our sparse approximation methods, leverage the low-rank representation of the system's input states rather than the inputs themselves.

Input states from sensor arrays or numerical discretization manifest with large dimension $N$ despite arising from a system of much lower rank $r$, where $r\ll N$. The proper orthogonal decomposition (POD) compresses high-dimensional data into a basis expansion consisting or $r$ basis POD modes. These POD modes $\bphi$ are the $r$ degrees of freedom in the system that span the spatial dynamics of system states $\bx\in\reals^N$ 
\begin{equation}
 \bx = \sum_{j=1}^r a_j\bphi_j(x)=\bPhi_r\ba. \label{eqn:spatialPOD}
\end{equation}
In the latter matrix form, the columns of $\bPhi$ are the POD modes and the vector $\ba\in\reals^r$ consists of $r$ POD coefficients that uniquely define any input state $\bx$.
Given $m$ states of a dynamical system stacked row-wise into a $N\times m$ data matrix $\bx=[\bx_1,\bx_2,\dots,\bx_m]$, the POD is efficiently computed from the singular value decomposition of $\bX$
\begin{equation}
\bX = \bPhi\bSigma\bV^T, \label{eqn:PODsvd}
\end{equation}
from which the {\em dominant} $r$ POD modes are retained in $\bPhi_r$. The POD coefficients for $\bx_k$ are $\ba = \bPhi_r^T \bx_k$. The strictly decreasing singular values $\sigma_j$ scale the POD coefficients $\bPhi_r^T\bX = \bSigma_r\bV_r^T$ and determine the truncation level $r$. In practice, $r$ is chosen so that the energy of the last dominant mode $\sigma_r/\sum_j \sigma_j$ is greater than some small threshold. 
\revision[R2:10]{The truncated POD basis is particularly suitable for low-rank approximation since the SVD is the explicit minimizer to the following optimization problem for a given target rank $r$,
	$$ \min_{\hat{\bX}}\|\bX - \hat{\bX}\|_F^2 \mbox{ subject to   rank}(\hat{\bX})=r,$$ 
or equivalently, $\hat{\bX} = \bPhi_r\bSigma_r\bV_r^T$. This is the well-known Eckart-Young Theorem \citep{eckart1936approximation}. The columns of $\bPhi_r$ span the column space of $\bX$, so they yield the best rank-$r$ least-squares approximation to the system $\bX$ in the Frobenius norm.
}{Because the modes are ordered by importance in the system in this manner, the  first $r$ POD modes are the best rank-$r$ least-squares approximation to the data $\bX$,
$$\bPhi_r=\argmin_{\tilde\bPhi_r}\|\bx-\tilde\bPhi_r\tilde\bPhi_r^T\bx\|_2.$$ }
For this reason the POD is a popular model reduction tool for complex flow dynamics and is broadly used across many fields in which it is alternatively known as principal component analysis (PCA) \citep{Pearson1901}, Karhunen-Lo{\`e}ve decomposition \citep{Loeve1955}, and Hotelling Transform \citep{Hotelling1933}.
\revision[R1:2]{POD modes are ordered by decreasing spectral energy (singular values), so the first few modes are sufficient for characterizing an intrinsically low-rank dynamical system. Lower energy modes can often be ignored.  This parsimonious representation of system states by a few dominant modes facilitates sparse classification within a library of POD mode sets from different classes, especially given significant variation between the different classes or in our case, flow environments.}{The POD is also an optimally sparse modal representation since the singular values concentrate energy in the first few modes, while remaining POD coefficients approach zero. This sparsity is important for sparsely approximating incoming states with a POD library spanning low-rank dynamical states from all regimes.}

\revision{\paragraph{Remark:} In subsequent sections, the truncation threshold $r$ is chosen so that the first $r$ normalized singular values capture 99\% of the energy in the system, or equivalently, sum to 0.99. This high energy threshold is suitable for noiseless data from physics simulations.
A more principled truncation method developed by \cite{gavish2014optimal} chooses an optimal truncation threshold based on noise level, singular value dropoff and data dimensions, which is more suitable for noisy data.}{}

\subsubsection{Library learning}
\begin{figure}
	\centering
	\begin{subfigure}[ht]{\textwidth} 
		\centering
		\begin{overpic}[width=0.8\textwidth]
			{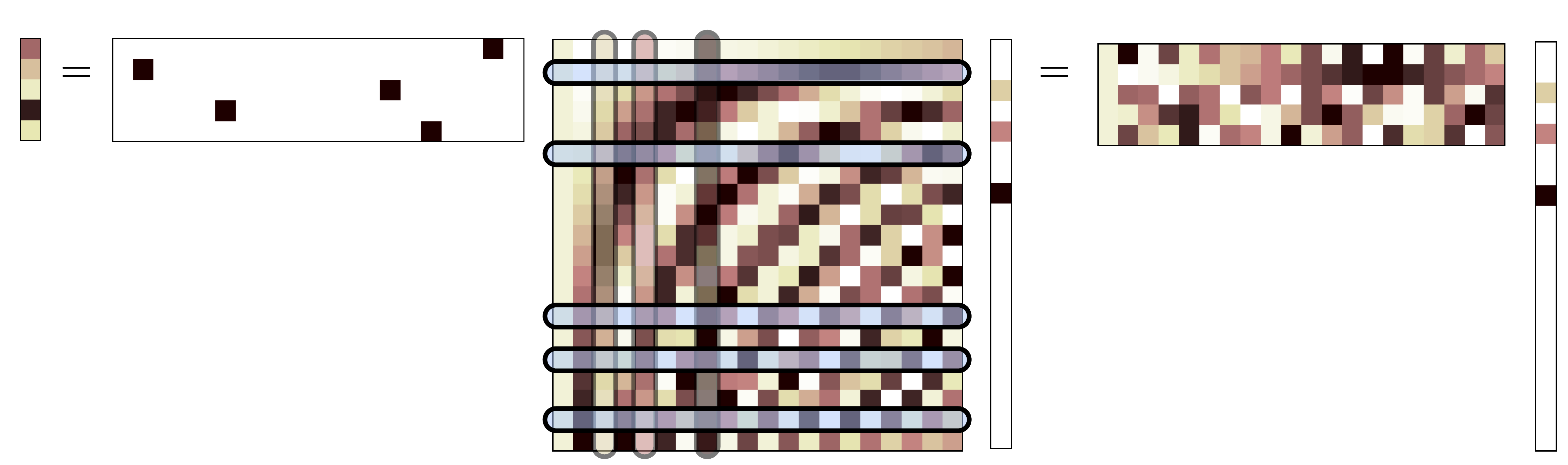}
			\put(1,28){$\by$}
			\put(20,28){$\bP$}
			\put(47,28){$\bPsi$}
			\put(81,28){$(\bP\bPsi)$}
			\put(98,28){$\ba$}
		\end{overpic}
		\caption{SRC}
	\end{subfigure}
	\caption{SRC yields sparse POD library coefficients $\ba$ from a few state measurements $\by=\bP\bx$ in an underdetermined linear system.  \label{fig:src_cs_comp}}
\end{figure}
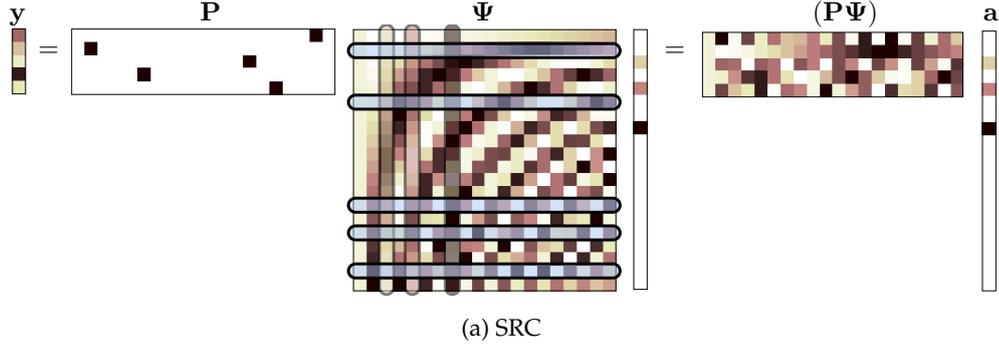
The supervised learning stage constructs a POD library from $c$ distinct environments expected to be encountered in flight. Data matrices of strain dynamics are collected through simulation for each flow environment $i$, which requires knowledge of current environment in the learning phase. The construction of truncated POD modes $\bPhi_{r_i}$ for each data matrix $\bX_i$ concludes the offline knowledge acquisition. The $c$ sets of POD modes are stacked into the library 
$$\bPsi  = \left[\bPhi_{r_1} ~|~ \bPhi_{r_2} ~|~ \dots ~|~ \bPhi_{r_c}\right].$$
New state measurements $\bx$ are then classified using POD library coefficients,
$$\ba = \left[\ba_{r_1}^T~~\ba_{r_2}^T\dots\ba_{r_c}^T\right]^T,$$
where its environment $i$ is identified by the set of POD modes that best approximate $\bx$ in a least-squares sense
\begin{equation}
\label{eqn:classifier}
 \mbox{environment}(\bx)=\argmin_i \| \bPhi_{r_i} \ba_{r_i}- \bx \|_2 .
\end{equation}
The classifier decision is therefore determined by the POD library coefficients, which cannot be obtained by the standard inner product with $\bx$ since $\bPsi$ is no longer unitary. Sparse library coefficients are sought so only the components of $\ba$ that correspond to the correct set of POD modes $\bPhi_{r_i}$ are nonzero. The library coefficients require a sparsity-promoting solution method that simultaneously solves for all $\ba_{r_i}$ at once, and we outline two such sparse approximation methods in what follows.

Library learning of low-rank {\em features} from data is well established in 
the computer science community.  More recently, the mathematical framework has migrated into the reduced order modeling community for characterizing parametrized PDEs~\citep{amsallem1,amsallem2,karen2,karen3,karen4,Sargsyan:2015}.  Thus libraries of ROM models that can be selected and/or interpolated  through measurement and classification.   Alternatively,  cluster-based reduced order models use a k-means clustering to build a Markov transition model between dynamical states~\citep{Kaiser2014jfm}.  Before these prototypical machine learning methods were considered for ROMs, it was already realized that parameter domains could be decomposed into subdomains and a local ROM/POD computed in each subdomain. \cite{sub1} used a partitioning based on a binary tree whereas \cite{sub2} used  a Voronoi Tessellation of the domain.  Such methods
were closely related to the work of \cite{sub3} where the data snapshots were 
partitioned into subsets and multiple reduced bases computed. The multiple bases were then recombined into a single basis, which is different than the library building techniques used more recently.  For a review of these domain partitioning strategies, please see Ref.~\cite{sub4}. 

\subsubsection{$L_1$ constrained approximation}
First we use the POD library and incoming full state measurements \revision{in signal $\bs$}{$\by$} to approximate library coefficients from the solution of the overdetermined linear system (Figure \ref{fig:src_cs_comp})
\begin{equation}
\label{eqn:overdetermined}
\bs\approx\bPsi\ba.
\end{equation}
The high-dimensional state $\bs$ (previously denoted $\bx$) is a linear combination of only one set of POD modes within the library and not the others, \ie, only a few adjacent library coefficients should be nonzero. 
Minimizing the number of nonzero components in a linear system (the $L_0$ norm of $\ba$) is a computationally intractable combinatorial search. However, relaxing the objective to $L_1$ norm minimization, where $\|\ba\|_1=\sum_k|a_k|$, results in the sparsest solution in many cases. The sparse solution is given by the optimization   
\begin{equation}
\label{eqn:src_objective}
 \ba = \argmin_{\tilde\ba} \| \tilde\ba \|_1 \mbox{ subject to } \| \bPsi\tilde\ba - \bs\|_2 < \epsilon, 
\end{equation}
where $\epsilon$ is a tunable error tolerance parameter. 
The performance is highly sensitive to the tolerance parameter - small values of $\epsilon$ risk approximating the least squares solution \eqref{eqn:leastsquares} and large $\epsilon$ risk library coefficients that are not sparse enough to select only one set of POD modes.
The identifier \eqref{eqn:classifier} relies on the subset selecting sparsity of $\ba$ that results from $L_1$ norm minimization.

\revision[R2:11]{On the other hand, the $L_2$ error minimizer given by the pseudo-inverse of $\bPsi$
	\begin{equation}	
	\label{eqn:leastsquares}
		\ba = \bPsi^\dagger \bs =
	\ba = \argmin_{\tilde\ba}  \| \bPsi\tilde\ba - \bs\|_2 ,
	\end{equation}
has the undesirable effect of weighing all available library vectors and distributing nonzero entries across all components of $\ba$, which confuses the classifier in \eqref{eqn:classifier}.}{We shall see later that the $L_2$ least-squares solution given by
\begin{equation}
 \ba = \argmin_{\tilde\ba}  \| \bPsi\tilde\ba - \by\|_2 .
\end{equation}
distributes weight across components of $\ba$, confusing the classifier \eqref{eqn:classifier}.}

\subsubsection{Sparse Representation for Classification (SRC) \label{sec:src_eim}}

A related idea, sparse representation for classification (SRC), was originally formulated by \cite{Wright:2009} for facial image recognition, and computational solution methods include convex optimization packages such as MATLAB's \verb+cvx+ \citep{cvx} and greedy algorithms such as CoSaMP \citep{cosamp}.
SRC acts on fewer state measurements than $R=\sum_{i=1}^c r_i$, the number of columns in $\bPsi$, resulting in an underdetermined system of equations with multiple solutions and no prescribed error tolerance. The measurement operator $\bP \in\reals^{p\times N}$ discretely samples $\bs$ at $p<R$ measurement locations that consists of $p$ rows of the $N\times N$ identity. Upon using the new state measurements $\by=\bP\bs$ and corresponding rows in the library $\bP\bPsi$, the system has multiple solutions $\tilde\ba$. The sparsest one is uniquely given by an optimization similar to \eqref{eqn:src_objective},
\begin{equation}
 \ba = \argmin_{\tilde\ba}  \| \tilde\ba \|_1 \mbox{ subject to } \by = \bP\bPsi\tilde\ba.
\end{equation} 
SRC is closely related to compressed sensing \citep{Donoho:2006,Candes:2006,Baraniuk:2007}. Compressed sensing is widely used for signal recovery from fewer random measurements than the signal's rank in the approximating basis. Perfect reconstruction requires $p=O(r\log(N/r))$ measurements that randomly encode the signal in a way that is incoherent with the approximating basis. This is alternatively known as the Restricted Isometry Property (RIP) that must be satisfied by the measurement matrix $\bP$. Later we show that $\bP$ can be optimized so that measurement locations are specially chosen to increase sparse classification accuracy. Furthermore, this can be done with fewer measurements than are required for reconstruction since classification is a milder objective.

In the reduced order modeling community, sparse sampling for state space reconstruction is not new.  Indeed, for nonlinear model reduction involving time-dependent PDEs, there are two critically enabling mathematical ideas:  (i) the rank-$r$ Galerkin projection of the dynamics onto POD modes, and (ii) the approximation of the nonlinearity and its inner products using gappy (sparse) POD sampling.
First introduced by Everson and Sirovich~\citep{gap1}, gappy POD~\citep{gap1,gap2,karni,Carlberg:2013}, and variants such as missing point estimation (MPE)~\citep{mpe}, {{``best points" method~\citep{patera}, empirical interpolation method (EIM)~\citep{eim}, and discrete empirical interpolation method (DEIM)~\citep{deim}, take advantage of sparse sampling strategies to project nonlinearities on the low-rank POD basis.  Specifically, the constraint $\by = \bP\bPsi\tilde\ba$ in (\ref{eqn:src_objective}) is identical with the gappy POD mathematical architecture where $\bP$ is prescribed by one of the above sampling methods.   In the ROM context, the projection is only on a specific, rank-$r$ POD basis so that $\bPsi$ is not a library of modes, thus allowing one to easily use standard $L_2$-based reconstruction.  In contrast, our sparse, $L_1$-based sampling strategy is used to select one of the many POD basis sets available for reconstruction.  Thus the objective of the sparse sampling in ROMs is quite different, yet relies on a similar mathematical framework.

\section{Fluid-Structure Interaction}\label{sec:fsi}

This section outlines our numerical approach for analyzing the fluid-structure coupling of the wing, and motivates sparse representation for classification in the frequency domain (SRCf).
The wing is modeled by classical elasticity theory ~\citep{GuentherLee:1988}.  As noted
previously, the governing equations model the deformation of the elastic body subject
to a function of the time-dependent chord loading $p(x,t)$.  
\revision[R2:12]{We define the spatial domain $x$ to represent the length of the wing chord that deforms in response to inertial-elastic forces. The range of $x$ is the closed interval $[0,2b]$ where $b$ is the half-chord length, $x=0$ represents the {\em leading edge} undergoing oscillatory actuation, and $x=2b$ is the stress and shear free {\em trailing edge}. These boundaries are consistent with biology - most insect wings are actuated by exoskeletal structure at their leading edge, while their trailing edge consisting primarily of thin membrane flutter at the behest of air loads without experiencing any inertial stress or shear forces. True wings deform in three dimensions, however we can approximate the deforming wing using two spatial dimensions $x$ and $z$ where $z(x)$ is the vertical displacement that depends on position along the chord. Assuming normal deflections and ignoring rotational effects, the elastic wing can be characterized by a forced linear Euler Bernoulli beam with spatially varying flexural stiffness $EI(x)$ and a zero deformation initial state that corresponds to a flat stationary wing. However, the strain data is collected after the deformation achieves a steady state, ignoring any initial transients. The governing equations and boundary conditions are thus given by:}{In addition to a governing
	partial differential equation, appropriate initial and boundary conditions are also required. The initial state and four boundary conditions (two at the leading edge and two at the 
	trailing edge) are specified as follows:}
\begin{subequations}
\label{eqn:beamboundary}
\begin{align}
 &\mu z_{tt} + (EI z_{xx})_{xx} = p(x,t), & &x\in(0,2b),t\ge 0 &\\
 & z(x,0) = 0 & &z_t(x,0) = 0 \\
 &z(0,t) = h(t) & &z_{xx}(2b,t) = 0  \\
 &z_x(0,t) = -\sin(\alpha(t)) &  &(EIz_{xx}(2b,t))_x = 0 
\end{align}
\end{subequations}
where $z(x,t)$ is the displacement of the elastic body along the length $x\in(0,2b)$, $\alpha(t)$
and $h(t)$ specify the
leading edge time dynamics for pitching and plunging respectively, and $p(x,t)$ is
the applied pressure loading that is computed from the Theodorsen model.  The
trailing edge boundary conditions specified at $x=2b$ model a shear-free and stress-free boundary.

\begin{figure}
	\centering
	\begin{overpic}[width=.32\textwidth]
	{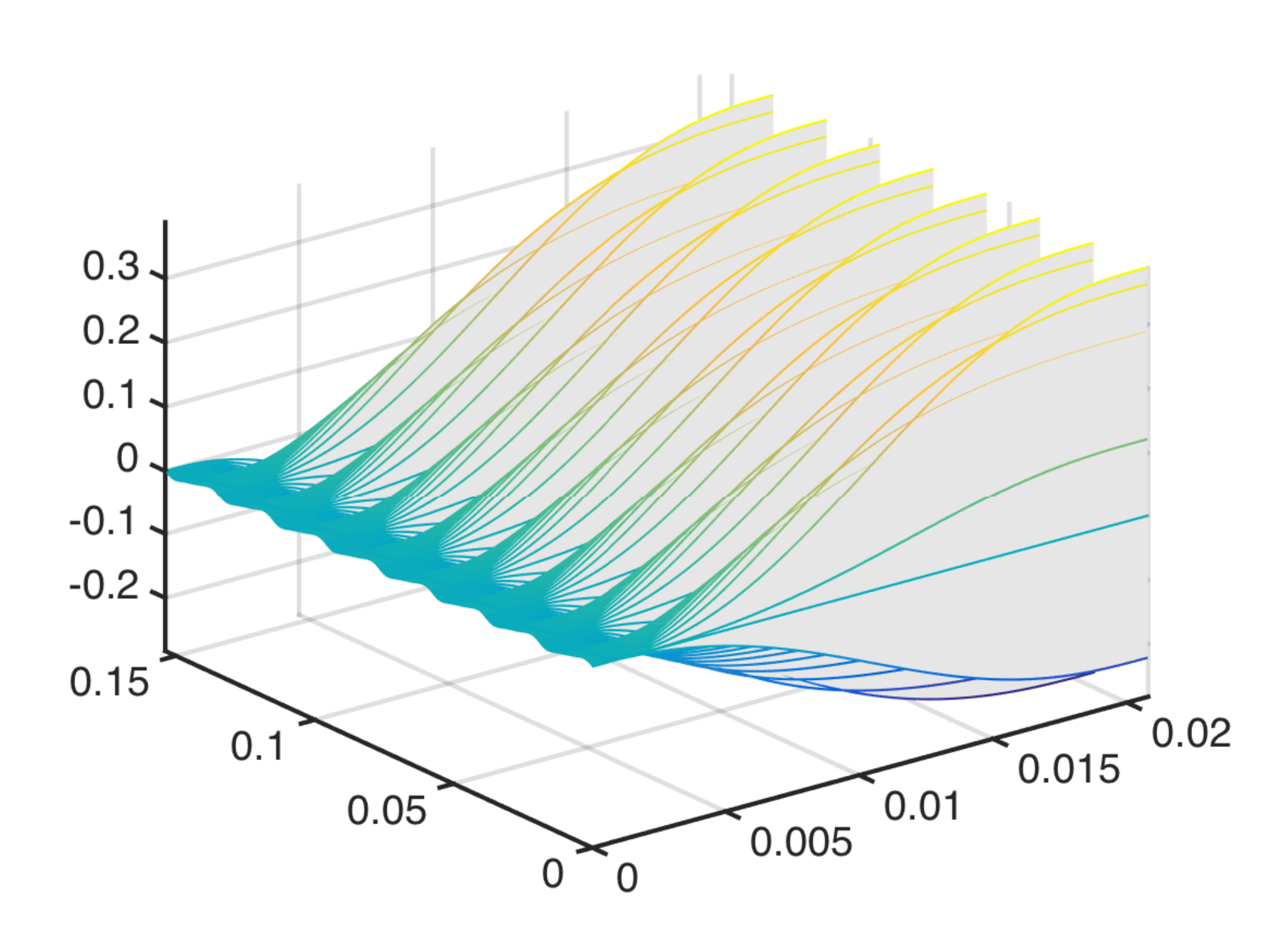}
	\put(35,70){Unforced}
	\put(0,40){\makebox(0,0){\rotatebox{90}{$s(x,t)$}}}
	\put(10,10){$t$[s]}
	\put(70,0){$x$[m]}
	\end{overpic}
	\begin{overpic}[width=.32\textwidth]
	{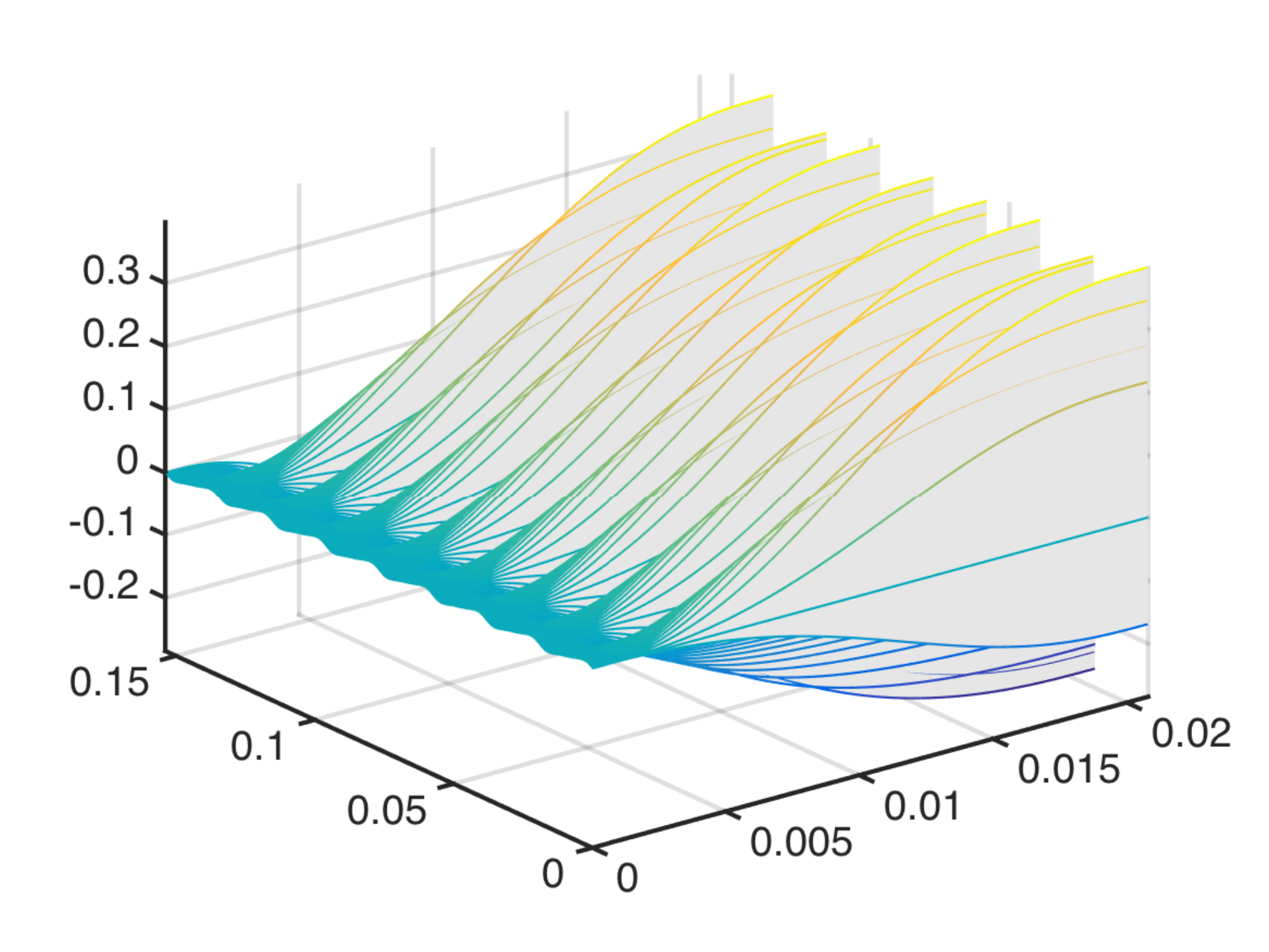}	
	\put(30,70){Forced, $f=5$ Hz}
	\put(10,10){$t$[s]}
	\put(70,0){$x$[m]}
	\end{overpic}
	\begin{overpic}[width=.32\textwidth]
	{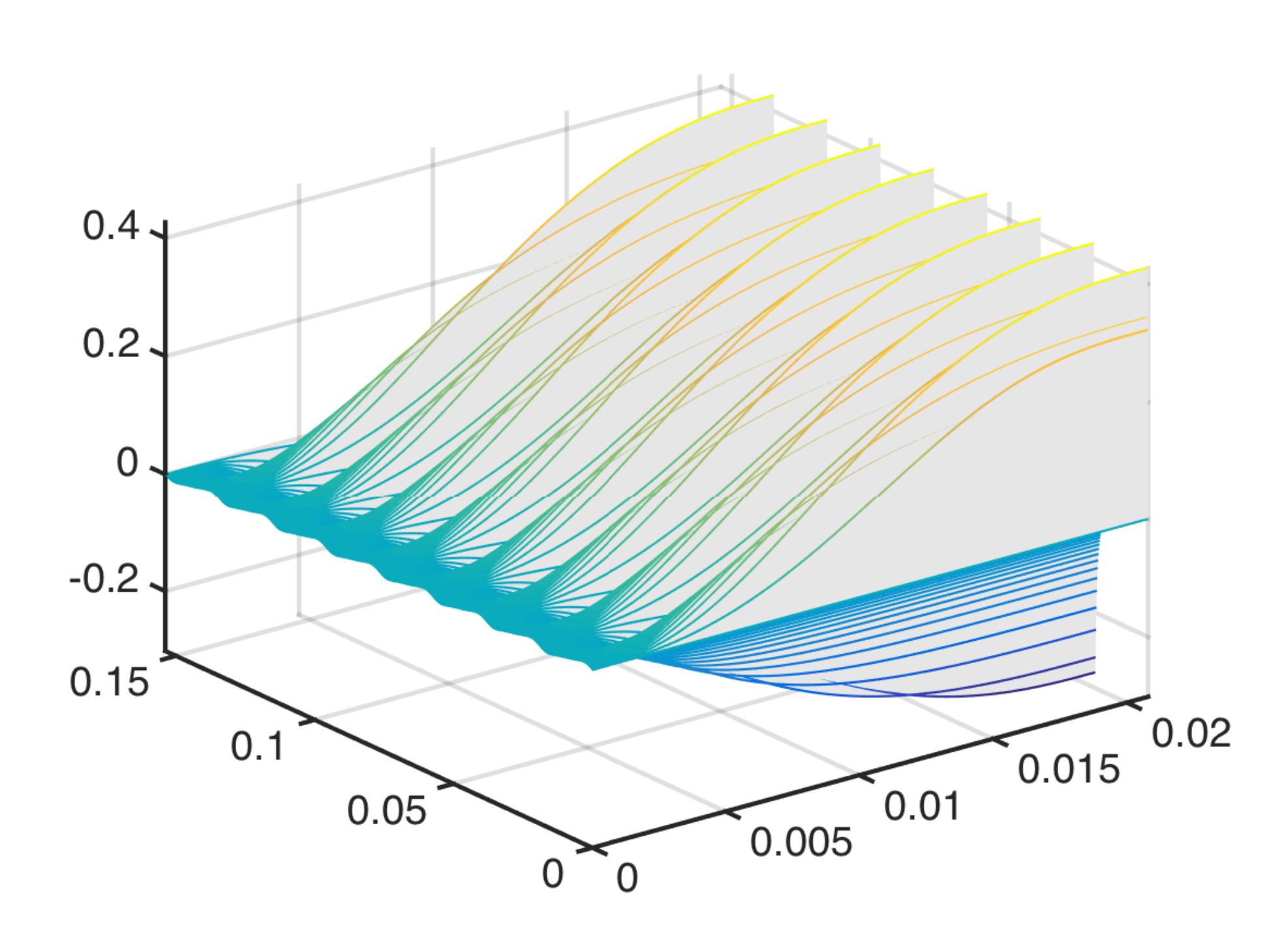}
	\put(30,70){Forced, $f=51$ Hz}
	\put(10,10){$t$[s]}
	\put(70,0){$x$[m]}
	\end{overpic}
	\caption{The strain dynamics for three different regimes, from left to right, (1) No gust disturbance, (2) low frequency gust disturbance $f=5$ Hz, (3) high frequency gust disturbance $f=51$ Hz. All three cases experience leading edge wing actuation at 26 Hz. Direct visualization of the strain dynamics offers no insight into the changing gust feedback between environments since inertial-elastic forces dominate over aerodynamic feedback forces.\label{fig:strainplots}}
\end{figure}
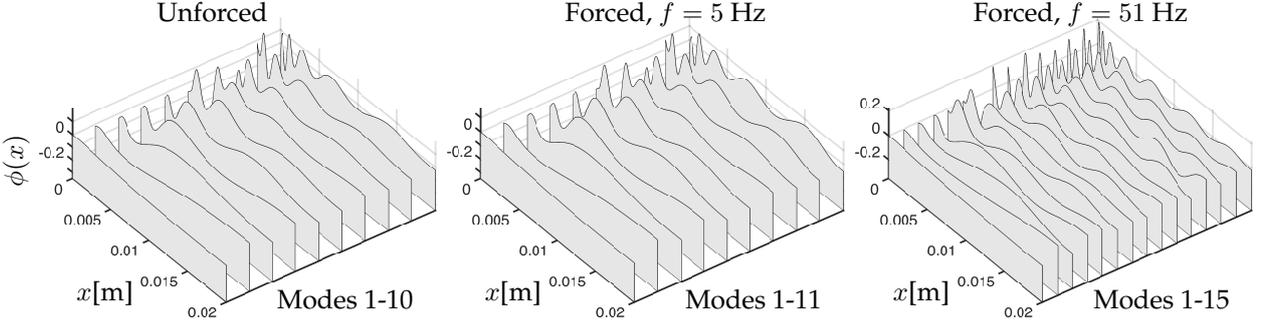
\begin{figure}
	\centering
	\begin{overpic}[width=.32\textwidth]
	{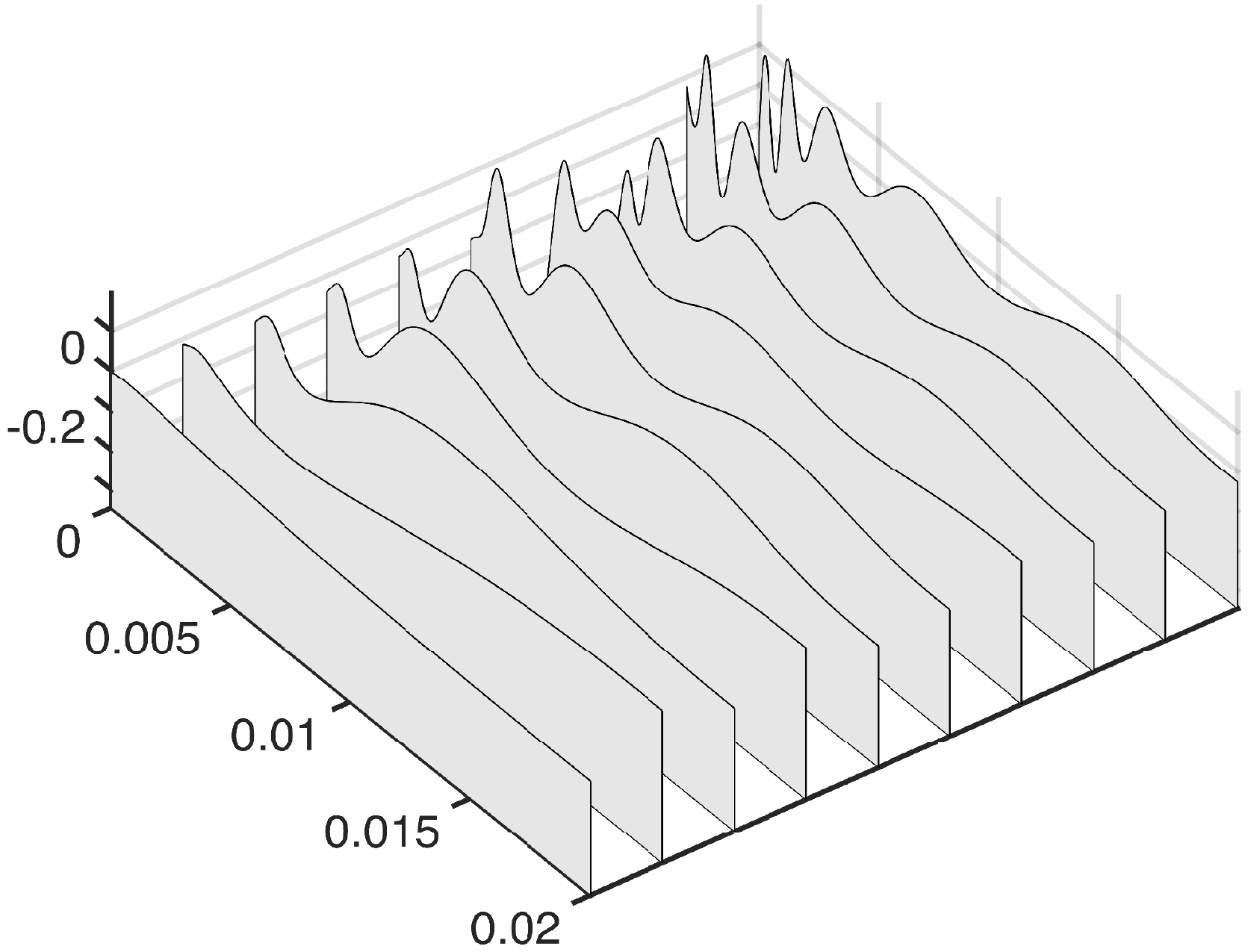}
	\put(30,75){Unforced}
	\put(-5,40){\makebox(0,0){\rotatebox{90}{$\phi(x)$}}}
	\put(10,5){$x$[m]}
	\put(60,3){Modes 1-10}
	\end{overpic}
	\begin{overpic}[width=.32\textwidth]
	{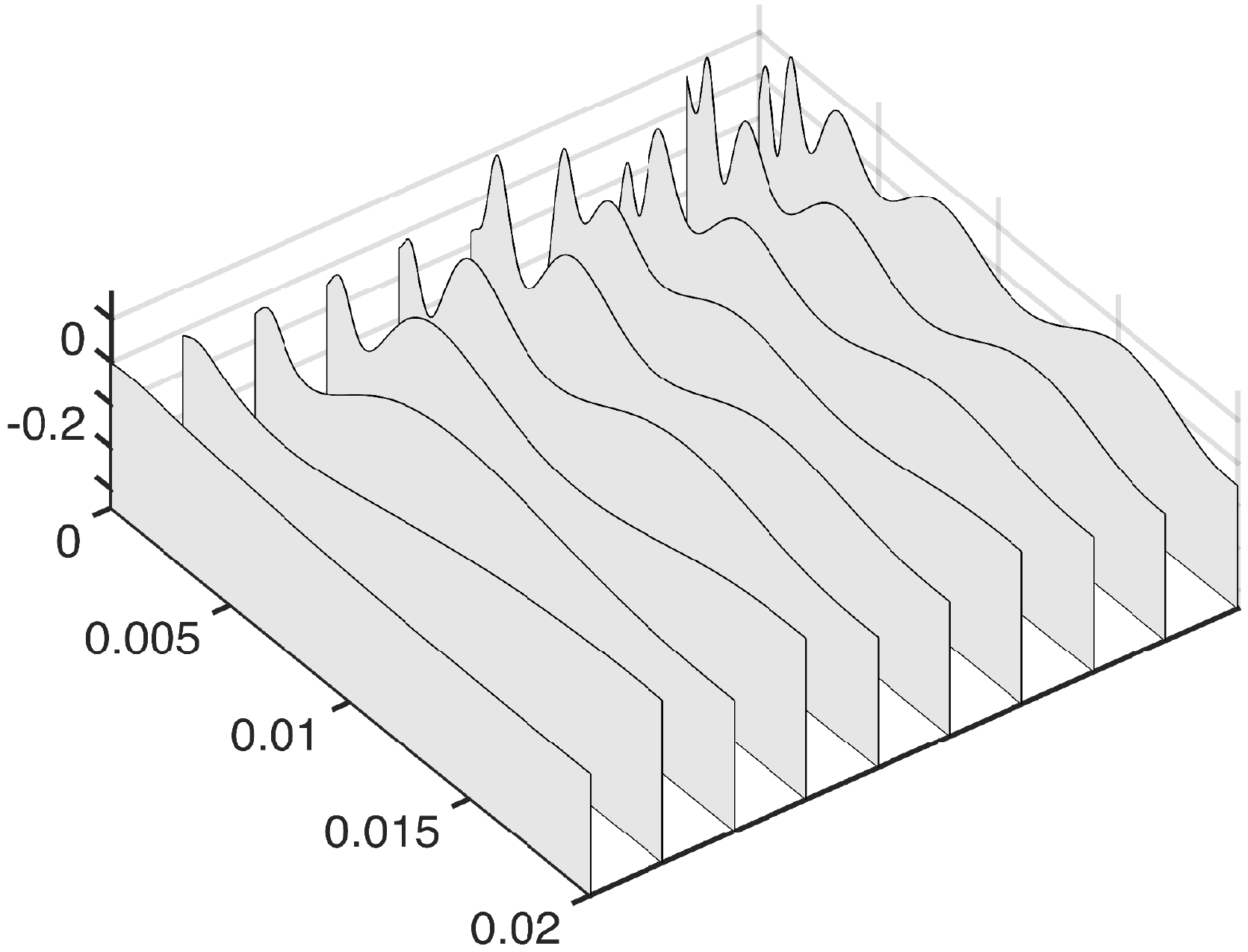}
	\put(30,75){Forced, $f=5$ Hz}
	\put(10,5){$x$[m]}
	\put(60,3){Modes 1-11}
	\end{overpic}
	\begin{overpic}[width=.32\textwidth]
	{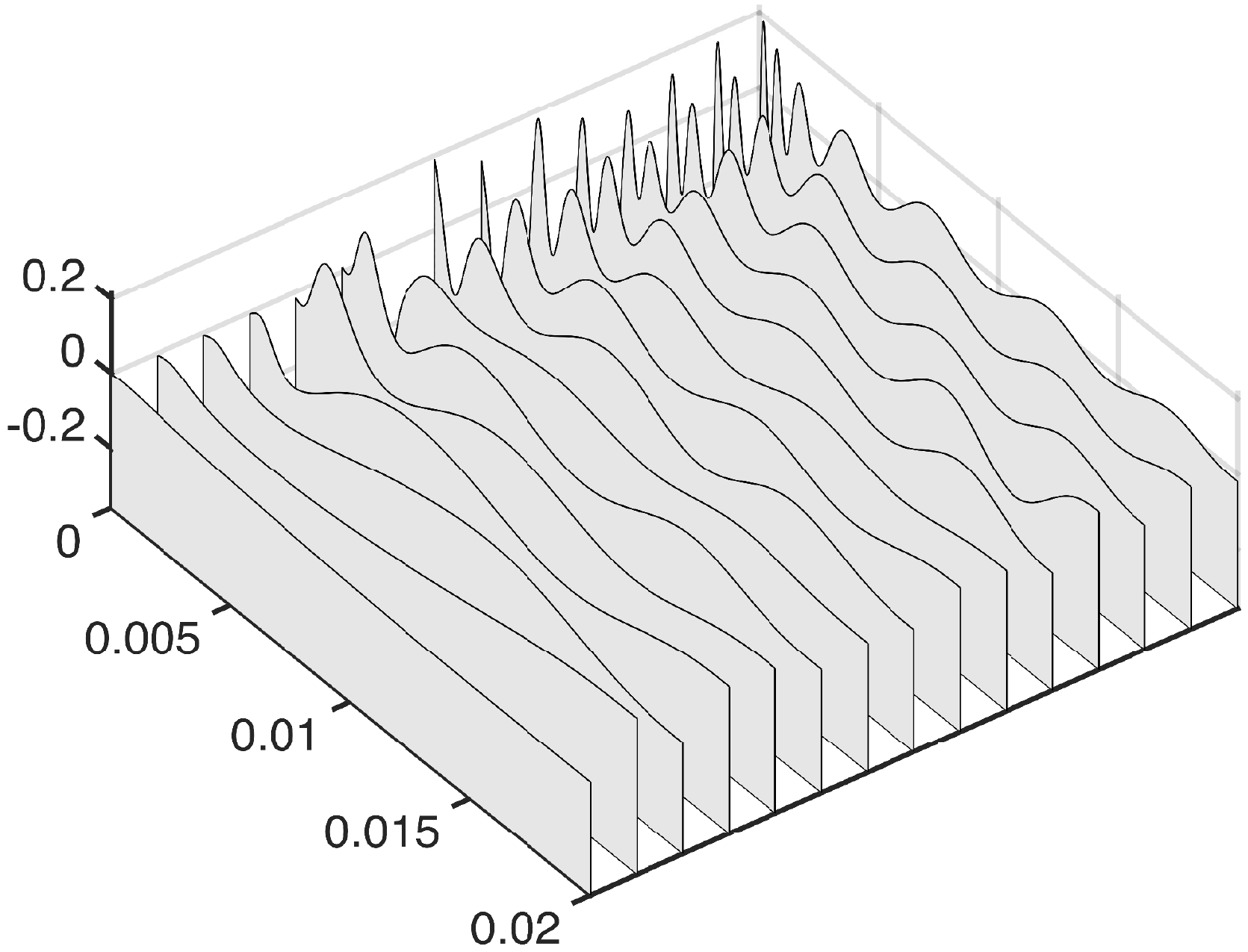}
	\put(30,75){Forced, $f=51$ Hz}
	\put(10,5){$x$[m]}
	\put(60,3){Modes 1-15}
	\end{overpic}
	\caption{The similarity of dominant spatial POD modes of wing strain between regimes 1 to 3 (Figure \ref{fig:strainplots}) suggests that spatial POD modes are not ideal for the SRC classification task. \label{fig:spatialmodes}}
\end{figure}
Numerical simulations use wing morphological parameters of the {\em Manduca sexta} hawkmoth, a representative species capable of hovering and other small-scale maneuvers that typify insect flight. The numerical model consists of two components that are weakly coupled as shown in Figure \ref{fig:airfoilwingcomp}: (1) elastic deformation strains and (2) unsteady aerodynamic loads. The deformation $z(x,t)$ is simulated with the hawkmoth pitching and plunging motion at 26 Hz at the leading edge. The {\em Manduca} wing's morphological parameters \citep{Combes:2002,Combes:2003} facilitate analysis of a small parameter space of a simplified 2D domain - the deforming wing chord along the $xz$ plane. \revision[R2:16] {Accordingly, the flexural stiffness distribution $EI(x)=10^{-5}e^{-150x}$ is determined by averaged experimental values from the actual hawkmoth in \cite{Combes:2002}, in which wings are shown to exhibit exponentially decreasing chordwise wing stiffness away from the leading edge.}{} An implicit finite-difference scheme is used to solve the dynamic 1D beam equation for the normal deformation $z(x,t)$ with spatially varying flexural stiffness to investigate the resulting unsteady fluid pressures and lift forces. Denoting the vector of transverse deflection at time $t_i$ as $\bz_{t_i}$, the strain vector $\bs$ is computed from the componentwise operation on $\bz$, 
\begin{equation}
 \label{eqn:strain} \bs(\bx,t_i) = \frac{\sqrt{ \Delta\bz_{t_i}^2 + \Delta x^2}-\sqrt{ \Delta\bz_{t_{i-1}}^2 +\Delta x^2}}{\sqrt{ \Delta\bz_{t_{i-1}}^2+\Delta x^2}},
\end{equation}
where $\Delta \bz_{t_i}$ denotes the spatial one-sided difference $\bz(x+\Delta x)-\bz(x)$ at time $t_i$. 

\begin{figure}
	\begin{overpic}[width=.32\textwidth]
		{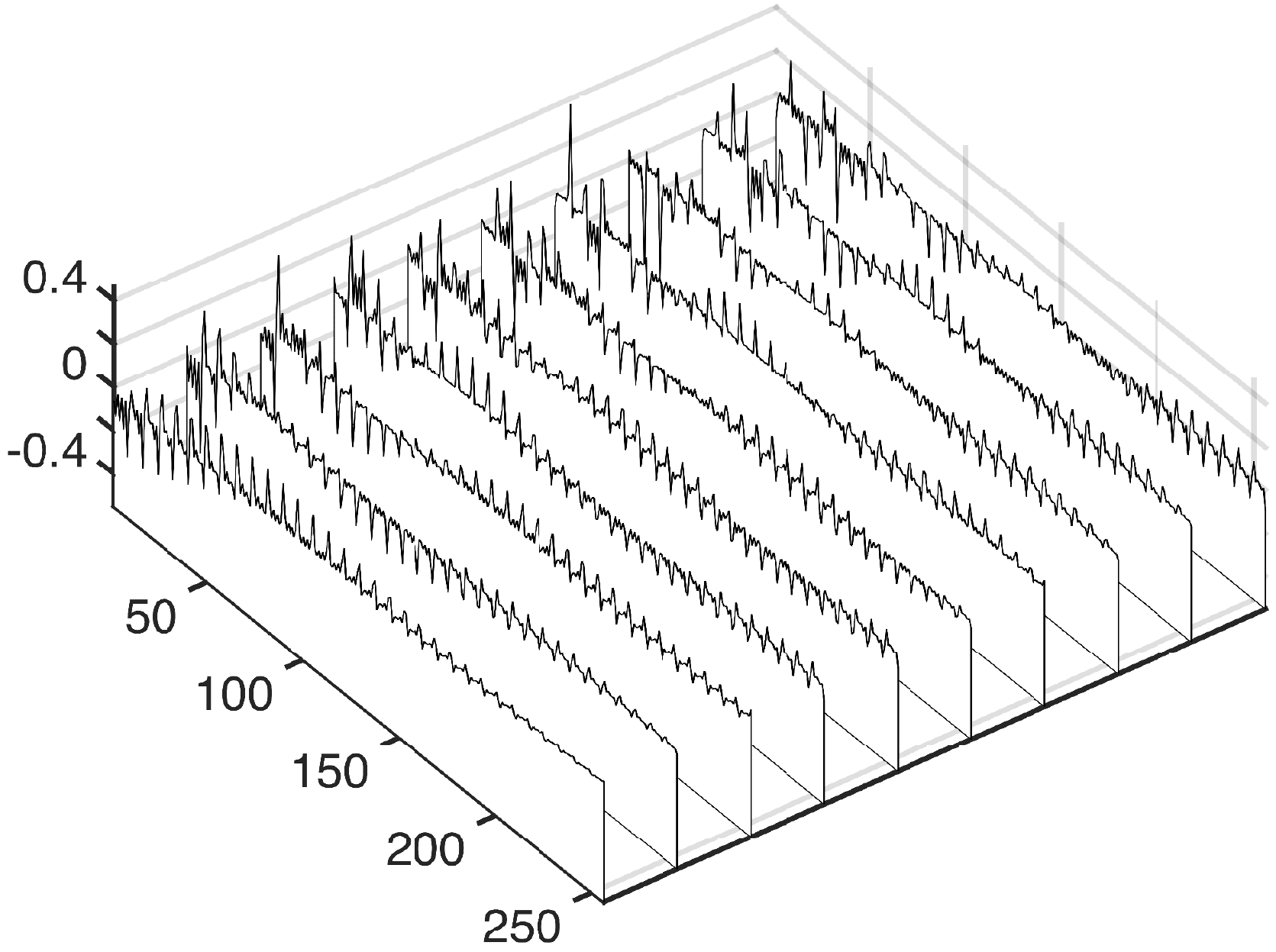}
		\put(10,10){$f$}
		\put(45,70){\color{RoyalBlue}$\hat{\bPhi}_{r_1}$}
		\put(60,3){Modes 1-11}
	\end{overpic}
	\begin{overpic}[width=.32\textwidth]
		{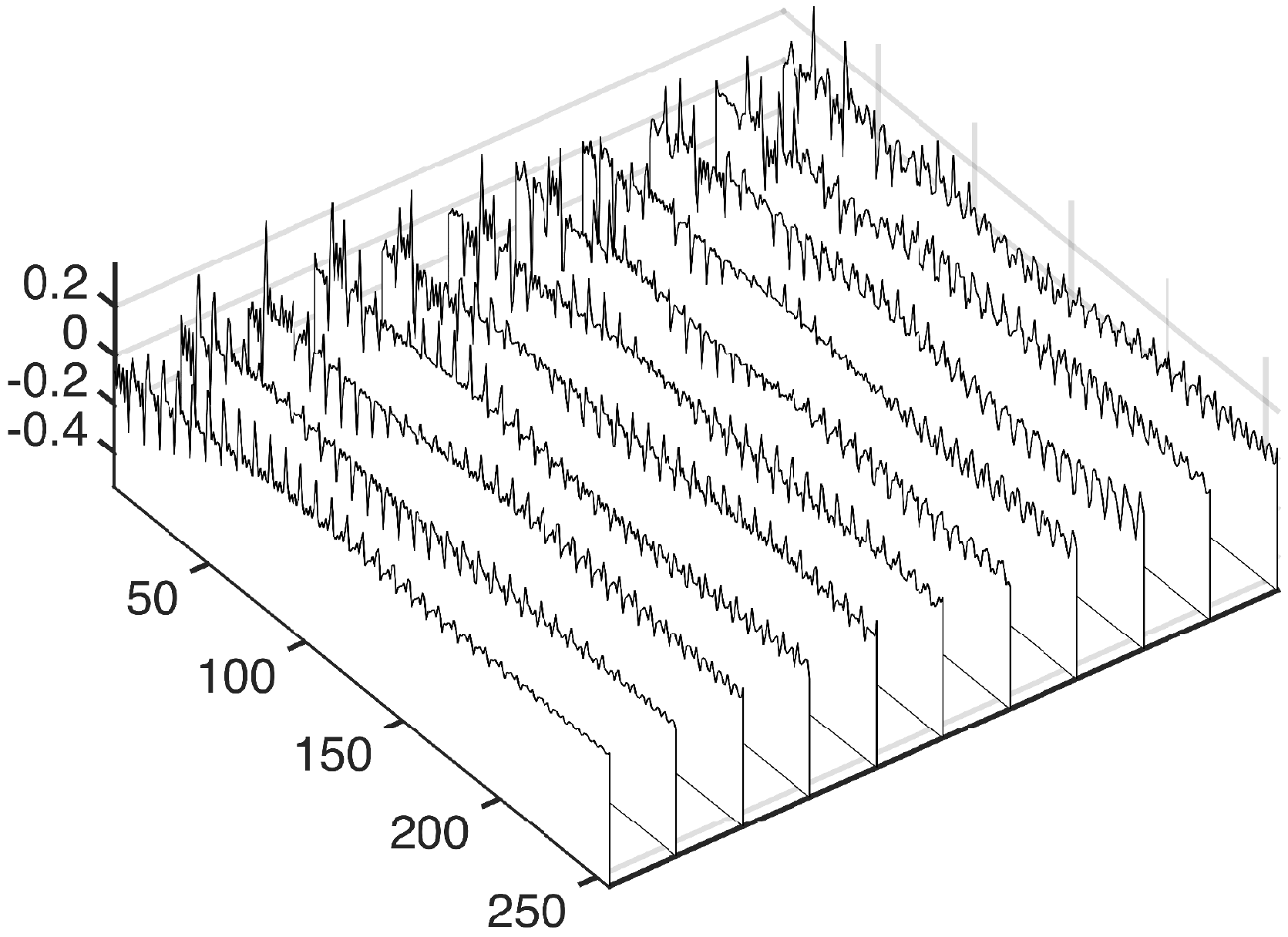}
		\put(10,10){$f$}
		\put(45,70){\color{BrickRed}$\hat{\bPhi}_{r_2}$}
		\put(60,3){Modes 1-11}
	\end{overpic}
	\begin{overpic}[width=.32\textwidth]
		{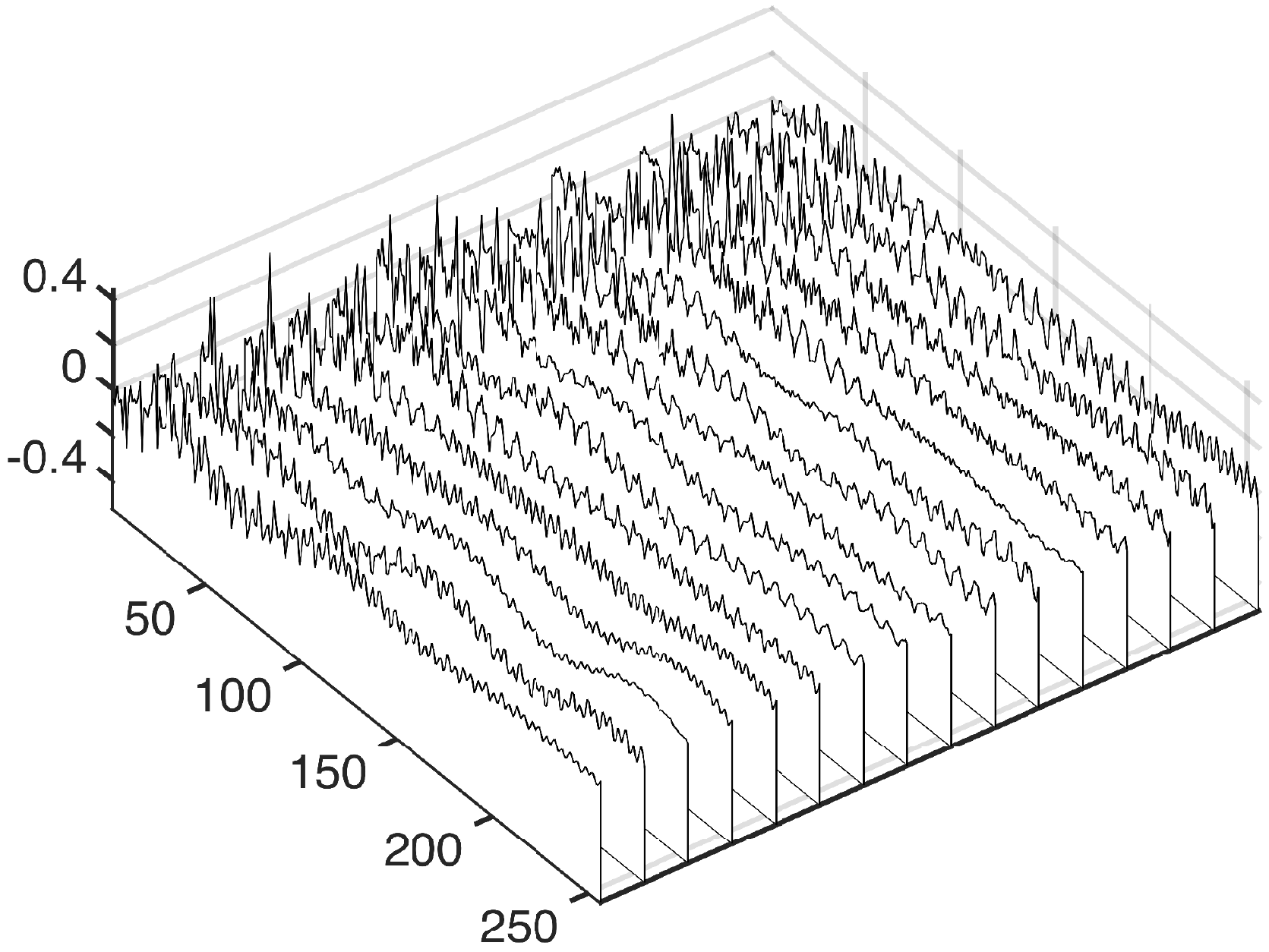}
		\put(10,10){$f$}
		\put(45,70){\color{ForestGreen}$\hat{\bPhi}_{r_3}$}
		\put(60,3){Modes 1-14}
	\end{overpic}
	\caption{Dominant POD modes of frequency content for regimes 1 to 3 (left to right) concatenated into an overcomplete library. This representation is ideal for classification because the frequency content differs markedly across the three regimes. \label{fig:fouriermodes}}
\end{figure}
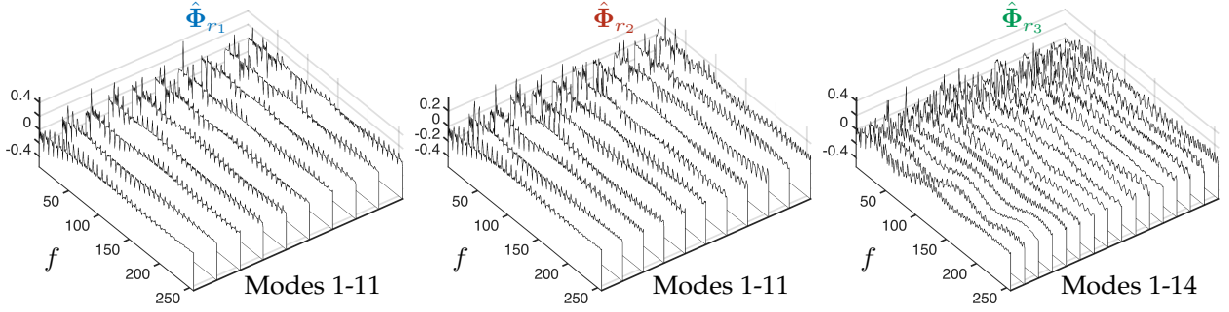
\revision{Meanwhile, the loads on the insect wing, $p(x,t;k)$, result from two sources -- the shed wake as well as external fluid disturbances. The first source of loads from the shed wake includes added masses and unsteady effects of the wing oscillation. These contributions are computed using the integral equation \eqref{eqn:pressure}, which expresses the local chordwise loads $p(x,t;k)$ as a function of $x$ and normal wing velocity $z_t(x,t)$. \cite{Abdo:2004} derived this expression explicitly as an intermediate step to Theodorsen's result for the space-averaged global pressure coefficient. However, the resulting changes in wing strain are small in magnitude and are characterized by the same dominant reduced frequency of wing oscillation at 26 Hz. }{We model the unsteady effect of the wing perturbing the surrounding fluid with Theodorsen's intermediate integral equation for the chordwise loads directly from the normal velocity of the wing $z_t$, explicitly stated in }
\begin{align}
\label{eqn:pressure}
p(x,t;k) &= \frac{2}{\pi}\mu U_\infty \sqrt{\frac{1-x}{1+x}}\int_{-1}^1 \frac{z_t(\td x,t)}{x-\td x}\sqrt{\frac{1+\td x}{1-\td x}} d\td x  + 
\frac{1}{\pi}\mu b \int_{-1}^{1} z_{tt}(\td x,t) L(x,\td x) d\td x \nonumber \\
&~~~~ + \frac{2}{\pi}\mu U_\infty[1-C(k)]\sqrt{\frac{1-x}{1+x}}\int_{-1}^1 z_t(\td x,t) \sqrt{\frac{1+\td x}{1-\td x}} d\td x, 
\end{align}
where 
$$L(x,\tilde{x}) = \log\frac{(x-\tilde{x})^2+(\sqrt{1-x^2}-\sqrt{1-\tilde{x}^2})^2} {(x-\tilde{x})^2+(\sqrt{1-x^2}+\sqrt{1-\tilde{x}^2})^2}.$$
\revision[R2:17,18] {The second source of chordwise loads are assumed to be external sinusoidal gusts characterized by different frequencies. We assume these gusts generate loads that resemble forces that would occur if the wing oscillates precisely at the characteristic frequencies of sinusoidal gust. Consequently, the loads resulting from external gusts are also computed using Theodorsen's model. This assumption approximates the aeroelastic effects of sinusoidal gust fields, since the forcing to the elastic beam includes these loads. This forced coupling between the elasticity model and the aerodynamic model addresses the absence of inertial effects in the Theodorsen framework. Specifically, we evaluate the following expressions for external loads, $p(x,t;k) = p_\alpha(x,t;k) + p_h(x,t;k) $, 
\begin{align}
\frac{p_\alpha(x,t;k)}{2b\mu \omega^2 \alpha_0 e^{i\omega t}} &= \left(\frac{U_\infty}{2}-\frac{i}{k}\right) \sqrt{1-x^2} + \frac{U_\infty}{2} x\sqrt{1-x^2} -\frac{C(k)}{k^2}\sqrt{\frac{1-x}{1+x}} - \frac{1}{2k}(1+2x+2C(k))\sqrt{\frac{1-x}{1+x}}, \quad\quad \\
\frac{p_h(x,t;k)}{2b\mu \omega^2 h_0 e^{i\omega t}} &= \sqrt{1-x^2} -\frac{iC(k)}{k}\sqrt{\frac{1-x}{1+x}},
\end{align}
given by \cite{PostelLeppert:1948} explicitly in their derivation of Theodorsen's lift coefficient \eqref{eqn:theodorsen_lift}. The resulting loads from the gusts represent varying flow environments, hence, we generate gusts characterized by a wide range of reduced frequencies. Since the model wing and base flow remain unchanged, this is accomplished by varying $f$, hence $k$, and holding free-stream velocity and chord length fixed.

First, wing strain dynamics are visualized for three environments $i=1,2,3$, all of which undergo the wing oscillation at 26 Hz. Environment 1 is unforced, and environments 2 and 3 undergo gust forcings of 5 Hz and 51 Hz, respectively. These frequencies correspond to reduced frequencies of $k \approx .03$ and  $k\approx 0.3$, the first of which is quasi-steady, and the latter represents unsteady flow. Figure \ref{fig:strainplots} displays the resulting strain dynamics which, by inspection, appear quite similar. The dominant spatial POD modes of strain, shown in \cref{fig:spatialmodes}, confirm that spatial modes remain largely unaffected by the changing forcing term. Indeed, without additional inertial or added masses in the model, the various forced dynamics are still governed by the intrinsic spatial modes. This poses a difficulty for data-driven classifiers based on spatial POD modes. Indeed, preliminary runs of SRC on strain snapshots from the three environments yield classification accuracies of 30\% on average, which is no better than random guesses. To remedy this, POD modes must incorporate the time dynamics of wing strain, as detailed in the following section. }{We generate aerodynamic loads from gust disturbances of varying frequencies $k$ using explicit expressions for chordwise loads $p(x,t) = p_\alpha(x,t) + p_h(x,t)$ derived by \cite{PostelLeppert:1948} from intermediate steps in Theodorsen's model not explicitly stated in \cite{Theodorsen:1935}: This expression yields the fluid loading resulting directly from the flapping which can be interpreted as aerodynamic feedback. However, this feedback only increases wing strain amplitude without introducing any interesting dynamics and is therefore considered negligible. Of more interest are the feedback dynamics arising from aerodynamic gust disturbances of different frequencies. Theodorsen's model is more commonly used as a function of the reduced frequency $k$ to obtain aerodynamic forces on the wing. The elastic beam simulations yield more interesting dynamics when aerodynamic loads excited by different $k$ are used as the forcing term $p$ in \eqref{eqn:beamboundary}. We assume here that aerodynamic disturbances occurring at different spatiotemporal scales than the wing flapping at 26 Hz will manifest as chordwise loads at differing frequencies from 26 Hz.}


\begin{figure}
	\hspace{1.5cm}
	\begin{minipage}[t]{0.39\textwidth} \vspace{0pt}
	\begin{tikzpicture}[scale=.95,framed,background rectangle/.style={draw=gray,rounded corners}]
	\draw[fill=red!20!white] (0,0) rectangle (0.3,4);
	\draw[fill=gray!20!white] (1,0) rectangle (3,4);
	\draw[fill=gray!20!white] (3.8-0.5,2) rectangle (4.1-0.5,4);
	\draw[fill=RoyalBlue!20!white] (4.25,0) rectangle (4.92,4);
	\draw[fill=red!20!white] (4.92,0) rectangle (5.59,4);
	\draw[fill=YellowGreen!20!white] (5.59,0) rectangle (6.26,4);
	\node at (0.15,3){$\hat\bs$};\node at (2,3) {$\hat\bPsi$};\node at (3.95-0.5,3) {$\ba$};
	\node at (0.6,3) {$\approx$};\node at (3.9,3) {$=$};\node at (6.5,3) {$\times$};
	\node at (4.25+.33,3) {\color{RoyalBlue}$\hat{\bPhi}_{r_1}$};
	\node at (4.92+.33,3) {\color{BrickRed}$\hat{\bPhi}_{r_2}$};
	\node at (5.59+.33,3) {\color{ForestGreen}$\hat{\bPhi}_{r_3}$};
	\node at (8.6,3){ };
	\node[align=center,font=\bfseries, yshift=2em] (title) 
	    at (current bounding box.north)
	    {Full state representation};
	\end{tikzpicture}
	\end{minipage}
	\begin{minipage}[t]{0.1\textwidth} \vspace{30pt}
	\begin{overpic}[width=\textwidth]
	{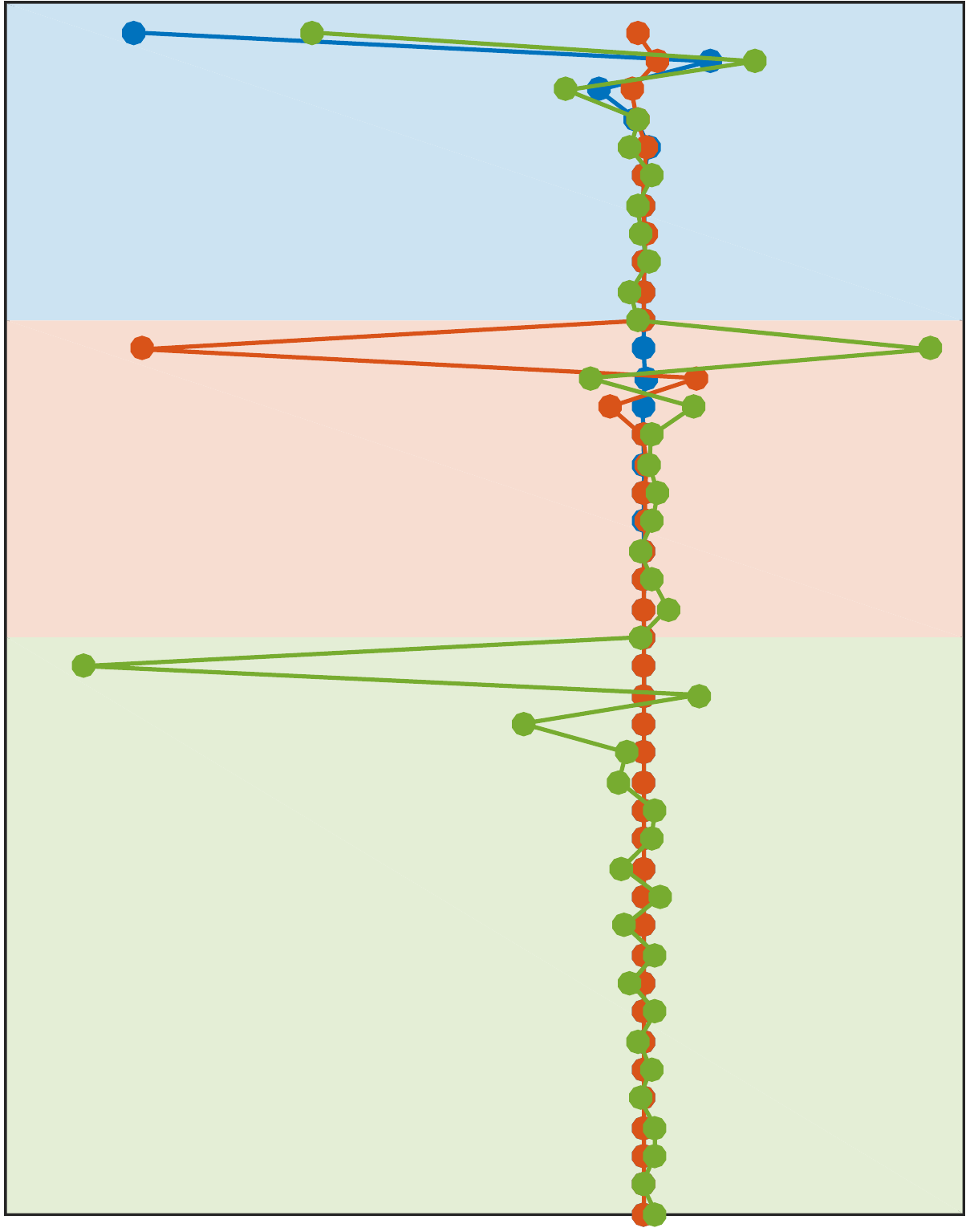}
	\put(25,50){$\ba$}
	\end{overpic}
	\end{minipage}
	~~~~
	\begin{minipage}[t]{0.1\textwidth} \vspace{0pt}
	\begin{tikzpicture}[scale=.95,framed,background rectangle/.style={draw=gray,rounded corners}]
	\draw[fill=red!20!white] (0.1,3) rectangle (0.4,4);
	\node at (0.25,2.75) {$\bP\hat\bs$~~$=$};
	\draw[fill=RoyalBlue!20!white] (1,3) rectangle (1.92-.25,4);
	\draw[fill=red!20!white] (1.92-.25,3) rectangle (2.59-.25,4);
	\draw[fill=YellowGreen!20!white] (2.59-.25,3) rectangle (3.26-.25,4);
	\draw[fill=gray!20!white] (4-0.75,2) rectangle (4.3-0.75,4);
	\node at (4-0.6,2.75) {$\ba$};
	\node at (2.1,2.75) {$\bP\hat\bPsi$~~~$\times$};
	\node at (0.7,3.5) {$=$};
	\node[align=center,font=\bfseries, yshift=2em] (title) 
	    at (current bounding box.north)
	    {SRCf};
	\end{tikzpicture}
	\end{minipage}
	\caption{SRCf promotes sparsity of the coefficients $\ba$ to help identify the correct originating environment $i$ using sparse frequency measurements $\bP\hat{\bs}$. The left plot contains 3 actual solution vectors. The blue line is the solution $\ba$ for a test vector from regime (1), red from (2), and green from (3). Each line correctly identifies its originating regime with its largest magnitude nonzero components.  \label{fig:SRCoverview}}
\end{figure}
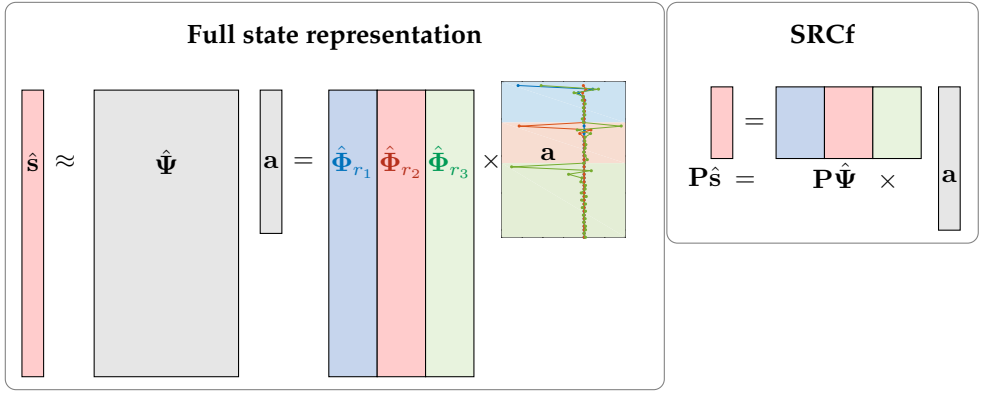

\section{Frequency domain representation of wing strain}

\revision[]{We introduce a sparse classification strategy that incorporates the time history of the strain signal through its frequency content. Indeed, insects and winged animals are thought to discriminate aerodynamic environments with only a few physical wing sensors that sample certain bandwidths of the feedback frequencies.}{Wing strain dynamics are simulated for environments $i=1,2,3$ with aerodynamic feedback frequencies of 5 Hz, 26 Hz and 51 Hz, respectively. Direct visualization of the dynamics shown in Figure \ref{fig:strainplots} offers no insight into the changing feedback dynamics, and by inspection the spatial mode shapes look similar. This presents a difficulty for the data-driven classifiers because POD library modes shown in Figure \ref{fig:spatialmodes} appear nearly identical for all three environments. Indeed, preliminary runs of SRC on strain snapshots from the three regimes yield an average classification accuracy of 30\% - no better than chance. This is not particularly surprising since inertial elastic forces govern the spatial mode shapes, and the inertial elastic parameters such as flexural stiffness, chord length, and flapping initial conditions at 26 Hz are fixed for all three regimes. The only parameter changing between regimes is $k$, the reduced frequency of the feedback from aerodynamic loads from a nonhomogeneous forcing term in the beam equation.}
%
%
\revision[R2:19]{Since gust forcing terms do not induce distinctive spatial modes, it is important to consider temporal strain dynamics, which are characterized by the jump discontinuities seen in \cref{fig:strainplots}. These discontinuities are distributed across many temporal POD modes, posing difficulties in obtaining a low-rank truncated POD. Hence, we transform the time dynamics into the frequency domain using the discrete cosine transform (DCT), and adapt the POD library and sparse classifier to classify the transformed strain dynamics. Thus the input to the sparse classifier, $\hat{\bs}$  (previously denoted $\bx$), can be expanded in terms of frequency domain POD modes,}{The POD library and sparse classifier are adapted to classify along strain sensor feedback frequencies instead of spatial snapshots of wing strain. This alternate POD basis expansion that spans the frequency content in the temporal direction is } 
\begin{equation}
 \bs(x,\bt) \xrightarrow{DCT} \hat\bs(x,\ff)  = \sum_{j=1}^r a_j\hat{\bphi}_j(f) = \hat\bPhi\ba. 
\end{equation}
This is distinct from the spatial POD modal expansion in \eqref{eqn:spatialPOD}. 
The numerical simulation for environment $i$ yields strain data $\bS_i$ for $m$ spatial gridpoints and $n$ timesteps. Recall that the standard POD is performed on a data matrix whose columns are time snapshots, that is, its columns span the spatial direction. For frequency domain analysis, the strain data matrices $\bS_i \in \reals^{n\times m}$ are adjusted so that its columns span time dynamics
$$\bS_i = [\bs(x_1,\bt) ~|~ \bs(x_2,\bt) ~|~ \dots ~|~ \bs(x_m,\bt)],$$
and the DCT applied to each column yields strain frequency content at every spatial location
$$\hat{\bS}_i = [\hat\bs(x_1,\ff) ~|~ \hat\bs(x_2,\ff) ~|~ \dots~|~ \hat\bs(x_m,\ff)].$$
Thus the procedure for building the POD library of frequency content remains unchanged
$$\hat\bS_i= \hat{\bPhi}_{r_i}\bSigma_{r_i} \bV_{r_i}^T,$$
where hatted notation indicates that POD modes span frequency content. Figure \ref{fig:fouriermodes} illustrates the dominant POD modes of the strain data's temporal frequency content for the three regimes. The frequency content reveals substantial differences between the aerodynamic regimes that facilitate robust, accurate classification.

\subsection{Classification in the frequency domain}
Given the frequency adjusted POD library, the SRC and compressed sensing procedures for classification remain identical. As before the POD modes $\hat\bPhi_{r_i}$ are stacked into the frequency POD library $\hat\bPsi$, and a new state is classified from the POD library coefficients. The \revision{input to the classifier}{signal to be classified} is the frequency content at one spatial location on the wing, $\hat\bs$. The $L_1$ constrained approximation for the full state is then given by 
\begin{equation}\label{eqn:src_objective2}
	\ba =\argmin_{\tilde\ba}  \| \tilde\ba \|_1 \mbox{ subject to } \| \hat{\bPsi}\tilde\ba - \hat{\bs}\|_2 < \epsilon.
\end{equation}
Note that the error tolerance $\epsilon$ must be chosen carefully to balance the trade-off between approximation and classification. Sampling the frequency content at a few limited frequency locations with $\bP$ yields the measurement vector $\bP\hat\bs$ used in SRC, with sparse solution coefficients given by $$ \ba =\argmin_{\tilde\ba}  \| \tilde\ba \|_1 \mbox{ subject to } (\bP\hat{\bPsi})\tilde\ba = \bP\hat{\bs}.$$ The SRC schemes are illustrated in Figure \ref{fig:SRCoverview}. As before, the strength of $\ba$'s components will determine the subset of regimes to which the strain frequency signal belongs. Because the classifier now operates in the frequency domain, SRC in the frequency domain is abbreviated as {\em SRCf}.

\subsection{\revtwo{Optimized frequency selection for SRCf}{Domain Samples}}
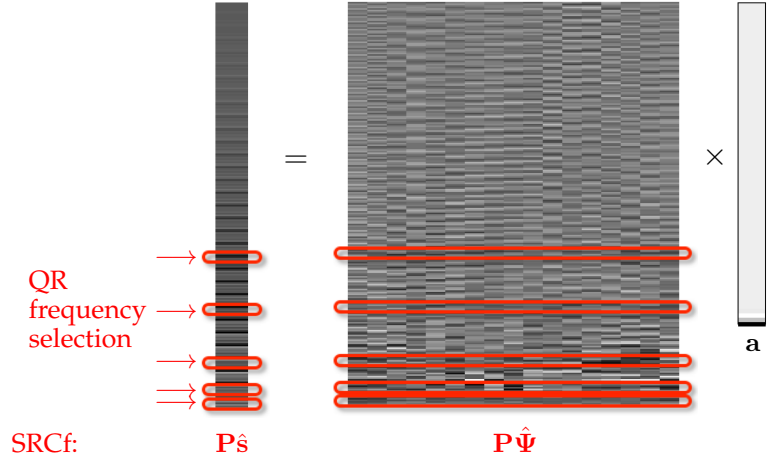
\begin{figure}
\centering
\begin{tabular}{m{0.1\textwidth} >{\centering}p{0.1\textwidth} >{\centering}p{0.3\textwidth} p{0.1\textwidth}}
 &\vspace{0pt} \begin{overpic}[width=.058\textwidth]{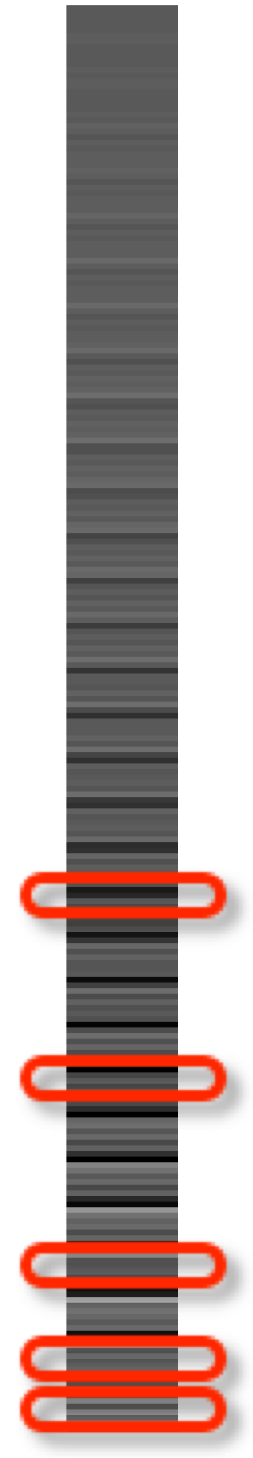}
\put(-40,31){\color{red}QR}
\put(-40,24){\color{red}frequency}
\put(-40,17){\color{red}selection}
\put(-10,2){$\color{red}\longrightarrow$} 
\put(-10,12){$\color{red}\longrightarrow$} 
\put(-10,24){$\color{red}\longrightarrow$} 
\put(-10,37){$\color{red}\longrightarrow$} 
\put(-10,5){$\color{red}\longrightarrow$} 
 \end{overpic} &
 \vspace{0pt} \begin{overpic}[width=.29\textwidth]{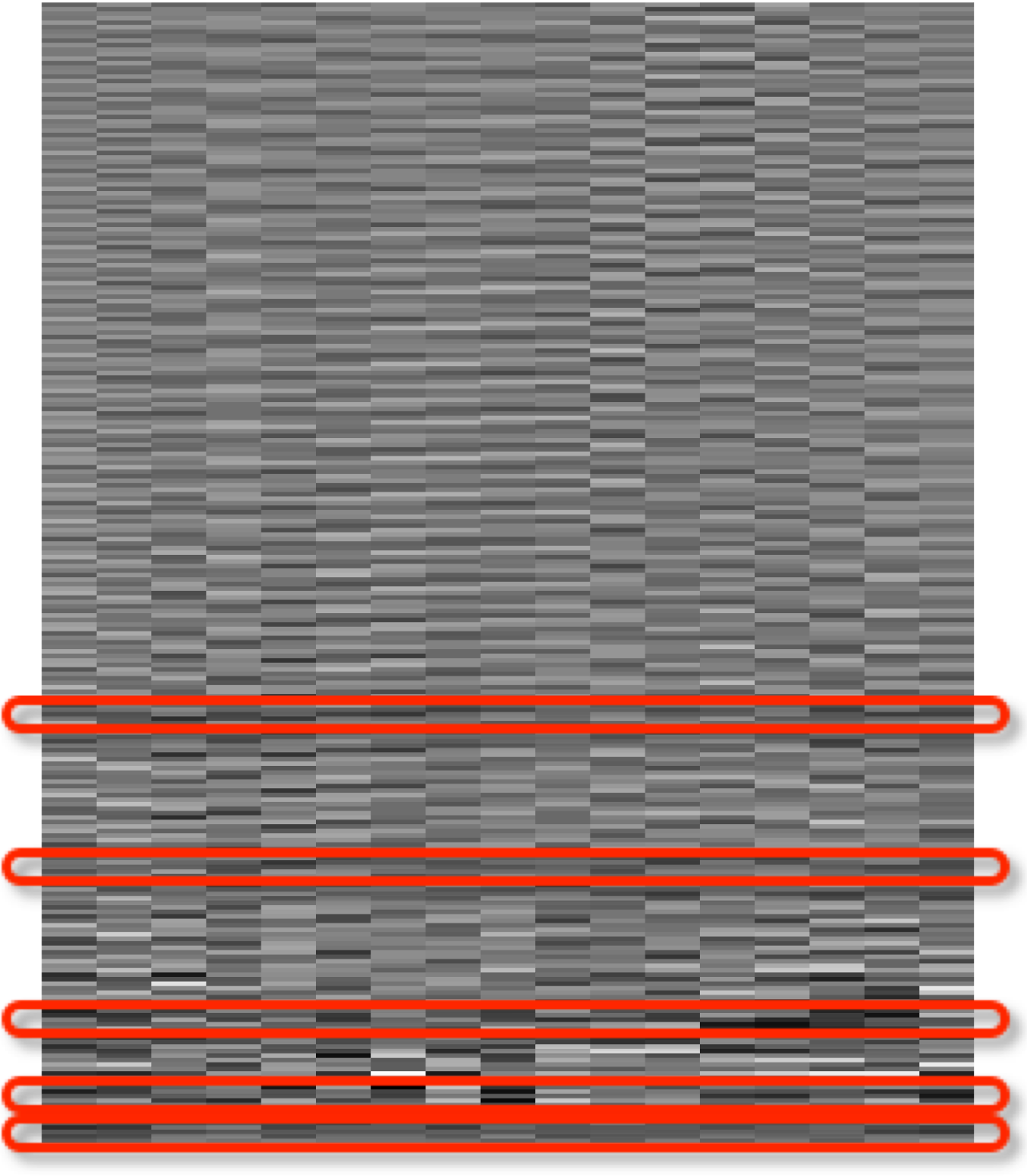} \put(-12,60){\large$=$}\end{overpic} &
 \vspace{0pt} \begin{overpic}[width=.03\textwidth]{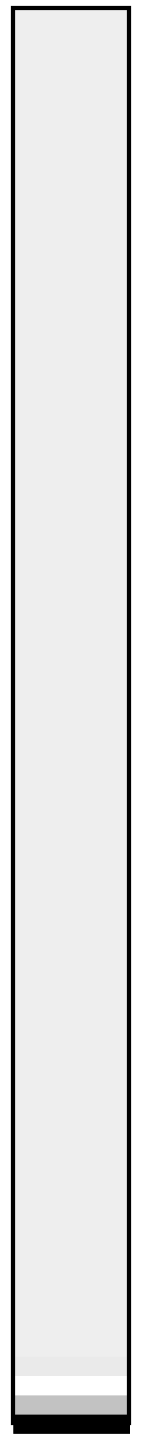}\put(-10,50){\large$\times$} \put(3,-7){$\ba$}\end{overpic} \\
\color{red}  SRCf: &$\color{red}\bP\hat\bs$ & $\color{red}\bP\hat\bPsi$ & 
\end{tabular}
\caption{This figure illustrates the action of the selection operator $\bP$ to the sparse approximation problem and converts the overdetermined SRCf into the underdetermined compressed SRCf problem.\label{fig:qdeimsrcf}}
\end{figure}

\revision[R1,R2]{
}{Compressed SRCf performance can be improved by intelligently choosing frequency measurement locations instead of random measurements. Fixed measurement locations should be located at selected frequencies that discriminate the training data by flow environment. This strategy is well-aligned with applications in closed-loop control and biology that optimize sensor locations to amplify desired feedback features. 

When data is available, state space measurement locations can be selected to preserve the structure of low-dimensional feature space. Existing work \citep{karni,gap2,eim,deim} has accomplished this by greedily selecting empirical interpolation points of features in the reduced basis. Flow identification requires the selected measurements to discriminate between all expected categories of flow environments in a reduced basis that preserves categorical variation in the data. This reduction is simply the POD of all available strain data from every flow environment at the cost of one additional SVD in the training phase
\begin{equation}
\hat\bTheta_r\bSigma_r\bV_r^T \approx [\hat\bS_1 ~|~ \hat\bS_2 ~|~ \dots \hat\bS_c].
\end{equation}
Given the truncated basis of dominant features $\hat\bTheta_r$, the interpolation points are the best $r$ discrete measurements $\bP\hat\bs$, where 
$$ \bP\hat\bs = \bP\hat\bTheta_r\ba,$$ 
necessary for full state reconstruction of $\hat\bs$ from its $r$ POD coefficients $\ba = (\bP\hat\bTheta_r)^{-1} (\bP\hat\bs)$. Since $\bP\hat\bTheta_r$ is a square matrix, optimizing measurement locations is equivalent to the optimization 
\begin{equation}
\bP = \argmin_{\tilde \bP} \|(\tilde \bP\hat\bTheta_r)^{-1}\|_2.
\end{equation}
This optimization finds the $\bP$ operator, or equivalently, the $r$ selected rows of $\hat\bTheta_r$ that preserve the dimension of the reduced space. The minimizer can be reliably and cheaply approximated by the QR factorization with column pivoting of $\hat\bTheta_r^T$,
\begin{equation}
\hat\bTheta_r^T\bP^T = \mathbf{QR}\bP^T,
\end{equation}
and outperforms existing sensor selection methods for gappy least squares reconstruction of a signal. This is shown in \cite{qdeim}, in which the authors refer to the QR column selection as Q-DEIM as it is closely related to DEIM.
The QR factorization pivots columns of $\hat\bTheta_r^T$ to contruct an $r$ column square submatrix of maximal rank, $\hat\bTheta_r^T\bP^T$. The transpose of the column pivoting is trivially the row selection $\bP\hat\bTheta_r$, and the selected columns are precisely the $r$ measurement locations in state space that best preserve the frequency space dynamics. The QR factorization additionally enjoys efficient implementation in many scientific computing packages, and a MATLAB implementation is included below:
Q-DEIM selects the same number of sensors as the truncation level $r$  of the reduced basis so the number of sensors can be varied as needed. For compressed SRCf, this $r$ is presumed to scale with the number of categories $c$.}%
Engineering and biological applications typically benefit from \revtwo[R2:1]{optimized measurement strategies}{} for decision-making and estimation tasks. 
Recall that $L_1$ regularized SRC permits sparse input sampling using, for instance, ROM methods like empirical interpolation methods (\cref{sec:src_eim}). EIM advocates measurement locations optimized for regression within the POD basis. Since {\em locations} correspond to the frequency bandwidths accessible for measurement, \revtwo[R2:1]{we refer to this optimized measurement selection as frequency selection.}{Accordingly, we designate this measurement selection as sensor placement as a nod to physical sensors that would be placed in real systems for control.}
\revtwo{This section briefly overviews existing methods for measurement selection, followed by our application of EIM for optimized frequency selection within POD modes of strain dynamics.}{}

\revtwo{\subsubsection{Background on measurement selection}
While empirical interpolation as measurement selection is a relatively new idea, measurement selection itself is a well-researched problem in signal processing, machine learning, design of experiment (DoE) and control. It is often designated as {\em sensor selection} since it typically informs sensor placement in spatial domains. Since sensor selection in large domains is a combinatorially hard search across possible sensors, a variety of greedy optimization strategies have been advocated that scale better with the search space. }{}
Geometric approaches in \cite{doe:hochbaum1985geom,doe:gonzalez2001geom} treat sensors as disks and attempt to cover the measurement space. If strong spatial correlations exist,  sensors can be placed to favor regions that experience greater {\em entropy} or variance \citep{doe:cressie1991entropy,doe:shewry1987entropy}. Meanwhile, \cite{doe:caselton1984info,doe:krause2008} greedily maximize {\em mutual information} of a sensor, which quantifies how much information a sensor contains about unused locations.  More general uncertainty quantification techniques \citep{doe:zhu2006uq,doe:zimmerman2006uq} advocate sampling specifically for parameter estimation. \cite{joshi2009sensor} tackle sensor selection with convex optimization and survey  related optimal experiment design and Bayesian approaches. In related work, \cite{BrBrPrKu:2013} develop an $L_1$-based convex optimization that operates directly on POD modes for facial image classification called sparse sensor placement optimization for classification (SSPOC). In addition, an overview of sensor selection for Gaussian processes can be found in \cite{doe:krause2008}. 

\subsubsection{Empirical Interpolation}
Sparse sampling for state space reconstruction uses EIM/DEIM points (as discussed in \cref{sec:src_eim}) that are well-conditioned for least-squares reconstruction in a single low-rank POD basis. To generalize these methods for $L_1$ classification in a library of modes, we first obtain a single low-rank POD basis for all categories of data.
%
For categorical data, a joint POD of all categories yields fewer modes than a concatenated library of POD modes from each consecutive category. Importantly, both are assumed to characterize the same active subspace or column space, since a joint POD compactly represents redundant features shared between categories in a POD library. Explicitly, we construct a joint POD using strain data from all flow categories,
	\begin{equation}
	[\hat\bS_1 ~|~ \hat\bS_2 ~|~ \dots \hat\bS_c] = \bTheta\bSigma\bV^T,
	\end{equation}
and truncate the left singular vectors to retain only the first $r$ dominant POD modes. This is advantageous for engineering the fewest measurements possible for sparse classification, since
$$ r < \sum_{i=1}^c r_i. $$
Subsequently, sparse library coefficients $\ba$ may be recovered from selected measurements of the full signal, where we denote the measurement selection operator by $\bP$
\begin{equation}
	(\bP\hat{\bPsi})\ba \approx \bP\hat{\bs}.
\end{equation}
This row selection operator, $\bP\in\reals^{r\times n}$, consists of rows of the $n\times n$ identity, so that
$$\bP_{j,I_j} = 1.$$
For the moment, we proceed as though the unknown state $\hat{\bs}$ is reconstructed (not classified) in $\bTheta_r$, the POD feature space spanning all classes. Since $\hat{\bs}$ is unknown, it cannot be recovered from its POD coefficients $\bTheta_r^T\hat{\bs}$.
%
Instead, the goal is to optimize $\bP$ for approximating $\hat\bs$ from $\by=\bP\hat{\bs}$ as follows
$$ \bP_\star = \argmin_\bP \| \hat\bs - \bTheta_r(\bP\bTheta_r)^{-1} \by \|_2. $$
DEIM \citep{deim} casts this optimization as a conditioning problem in the spectral norm,
\begin{equation}
\label{eqn:volume_opt}
	\bP_\star = \argmin_\bP\|(\bP\bTheta_r)^{-1}\|_2,
\end{equation}
where the attained minimum is denoted $\gamma_\star = \|(\bP_\star\bTheta_r)^{-1}\|_2.$
Equation \eqref{eqn:volume_opt} also maximizes the matrix volume of the product $\bP\bTheta_r$, which is the absolute value of its determinant. A na\"{i}ve brute-force search across all combinatorial $n \choose r$ row selections (measurements) of $\bTheta_r$  is computationally intractable and would involve evaluating determinants for each selection. Empirical interpolation methods use greedy procedures to bypass this combinatorial search.
DEIM iteratively selects measurements one at a time by locating them at maxima of successive residuals from approximation with previously chosen measurements. It is extensively used since it is computationally cheap and has established upper bounds for the minimizer $\mathbf{\gamma}$. 

\cite{qdeim} demonstrate an even better choice for $\bP$ given by the column pivoted QR factorization of $\bTheta_r^T$ called {\em Q-DEIM}. QR column pivoting has been a pioneering workhorse for the solution of underdetermined linear systems ever since its introduction by \cite{businger1965qr}. Its utility in least-squares polynomial approximation has led to related work in finding near-optimal Fekete interpolation points from polynomial Vandermonde matrices \citep{sommariva2009fekete} and interpolation points in weighted polynomials \citep{seshadri2016qr}. This procedure is designated {\em QR selection} to distinguish sampling for sparse approximation in a POD library from Q-DEIM sampling for state reconstruction in ROMs.

The rest of this discussion closely follows \cite{qdeim}, in which the pivoted QR factorization of $\bTheta_r^T$ is shown to be a near-optimal solution of \eqref{eqn:volume_opt}. The QR factorization expresses $\bTheta_r^T$ as the product of an orthonormal matrix $\bQ$ and an upper-triangular matrix $\bR$. The column pivoted QR rearranges $r$ columns of $\bTheta_r^T$ so that its first $r$ columns can be expressed as a product of $\bQ$ and $\bR_1$, the leading square $r\times r$ submatrix of $\bR$, such that
\begin{equation}
\bTheta_r^T \mathbf{\Pi} = \mathbf{QR}_1.
\end{equation}
QR selection essentially pivots measurement space to construct a special diagonally dominant $\bR_1$ with a lower condition number than would have been possible without column pivoting. The resulting row selection is then given by
 $$\bP = \mathbf\Pi^T.$$
Construction of a diagonally dominant $\bR_1$ is beneficial because the spectral norm depends entirely on $\bR_1$, 
 $$ \gamma_{qr} = \|(\bP\bTheta_r)^{-1} \|_2 = \|\mathbf{R}_1^{-1}\mathbf{Q}^T\|_2 = \|\mathbf{R}_1^{-1}\|_2 = \frac{1}{\sigma_{\min}(\mathbf{R}_1)}.$$
Finally, \cite{qdeim} demonstrate the improvement of pivoted QR selection over the leading method, DEIM, in theory and in practice. They derive upper bounds for $\gamma_{qr}$ that are smaller than $\gamma_{deim}$. Moreover, in practice, QR selection values for $\gamma$ fall well below that of the DEIM procedure. Indeed, the authors demonstrate empirically that $\gamma_{qr}$ often satisfy the conjectured optimal upper bound $\gamma_\star$, while that of DEIM surpasses it. Thus, QR selection is near-optimal in the matrix volume maximizing criterion.

\subsubsection{QR Software Implementation}

Software implementations of QR are readily available in most scientific computing packages, including LAPACK, ScaLAPACK, and MATLAB. Most subroutines implement Businger-Golub pivoting using Householder projections as detailed in \cite{businger1965qr}, adapting the procedure as necessary to deal with rank-deficiency. However, any of the procedures may be used in this setting since we only consider matrices of full rank. After each successive orthogonal projection step, this method successively selects the next column with maximal 2-norm as the next pivot, with the effect of increasing the condition number of the column pivoted target matrix. Below is MATLAB code for constructing the QR selection operator $\bP_{qr}$, given $\bTheta_r\in\reals^{n\times r}$ and truncation level $r$:
\begin{lstlisting}
	%% MATLAB code for given Theta_r matrix
	
	n = size(Theta,1); % dimension of state space
	Theta_r = Theta(:,1:r); % first r POD modes
	
	% Pivoted QR factorization of POD modes transposed
	[Q,R,pivot] = qr(Theta_r','vector');
	
	% QR row selection indices
	ind = pivot(1:r);
	
	% Form row selection operator 
	I = eye(n);
	P_qr = I(ind,:);
\end{lstlisting}
More generally, QR measurement selection can easily be adapted to incorporate measurement location constraints as would be commonly encountered in engineering applications. Undesirable measurement locations from a practical standpoint can simply be omitted from the QR pivoting algorithm by omitting the appropriate rows (measurements) from the input $\bTheta_r$. More complex preferences can be implemented by multiplying the input modes by application-specific weight matrices; this is the focus of ongoing work.

\section{Results}

In this section sparse approximation simulations are designed to detect flow-induced strain patterns across various flow environments, frequency samples and spatial correlations. Importantly, the advantages of sparse over conventional least squares approximation and trained \revision[]{QR row selections}{Q-DEIM samples} over random samples in SRCf are demonstrated.
These patterns are of particular importance in engineering applications that require collection of strain data optimized for environment detection. These tests also stress the importance of sensor noise in these applications by demonstrating robust sparse approximation performance over a range of increasing noise levels in the strain dynamics.

In this section, machine learning terminology is used to describe the data. Single location frequency content are called observations or signals, to disambiguate from system states that may refer to snapshots in time. Observations to be classified, $\hat{\bs}$ or $\by=\bP\hat{\bs}$, are called inputs to the classifier. 
As a pre-processing step, noise is added to each column of the untransformed input observations in the following order.
\begin{enumerate}
\item Normalize each column so that $\|\bs(x_j,\bt)\|_2=1$ for all $j=1\rightarrow m$ gridpoints,
\item Add zero-mean Gaussian noise so $\bS = \bS+\xi$, $\xi\sim N(0,\eta^2)$, and
\item Transform $\bS$ into frequency domain $\hat\bS$: $\hat\bs(x_j,\ff) = DCT[\bs(x,\bt)]$.
\end{enumerate}
Thus sensor noise is added to the untransformed normalized signal in the space-time dimension as would be the case in engineering applications, and classifier inputs are normalized to facilitate comparing the effect of noise across different environmental regimes. Note that we do not normalize or add noise to the transformed signal $\hat\bs$ in the frequency domain since wing strain measurements are noisy when collected (sensor noise), not when processed via the frequency domain (process noise). Table~\ref{tab:summary} provides a summary of the numerical experiments performed and the figures produced for this section.

\begin{table*}[t]
	\centering
	\scalebox{1}{
		\begin{tabular}{l l l l l l l} 
			\hline 			\hline
			\multicolumn{1}{l}{\bf Method}
			& \multicolumn{1}{l}{\bf Figure}
			& \multicolumn{1}{l}{\parbox{2.4cm}{\bf Input \\dimension}}
			& \multicolumn{1}{l}{\parbox{2.4cm}{\bf Sensors}}
			& \multicolumn{1}{l}{\parbox{2.4cm}{\bf Gust \\frequencies} }	
			& \multicolumn{1}{l}{ \parbox{2.45cm}{\bf Number of \\ library modes}}						
			\\
			\cmidrule(r){1-6}
			
			\multirow{1}{*}{\rotatebox[origin=c]{0}{ \parbox{2.8cm}{$L_1$ approximation}  }} 
			& 8, \S 5.1	& 256 &  --  &  $\varnothing$,5,51  &  36 \\ 
			\hline			
	
			\multirow{1}{*}{\rotatebox[origin=c]{0}{ \parbox{2.8cm}{Least squares}  }} 
			& 9, \S 5.2	& 256 & --  &  $\varnothing$,5,51 &  36 \\ 
			\hline		
						
			\multirow{2}{*}{\rotatebox[origin=c]{0}{ \parbox{2.8cm}{ SRCf} }} 
			&  10, \S 5.3	& 5 &  QR   &  $\varnothing$,25,50,75,100  & 75  \\ 
			&  11, \S 5.3	& 10 &  QR  & $\varnothing$,25,50,75,100  &  75   \\ 		
			&  12, \S 5.4	&  35 & Random  & $\varnothing$,25,50,75,100  &  75   \\ 

			\hline \hline
		\end{tabular}
 	}
	\caption{Summary of numerical experiments and sparse classifiers, where $\varnothing$ refers to the flow environment with no gust forcing. All flow environments exhibit the dominant 26 Hz frequency of wing oscillation. Furthermore, the number of gust frequencies equals the total number of classes considered.\label{tab:summary}}
\end{table*}


\subsection{Feedback identification from full state} \label{ssec:srcf_regime}

We demonstrate sparse flow feedback identification between the three learned flow environments in \cref{sec:fsi}, and test it on randomly selected subsets of strain dynamics from each flow environment. The environments are chosen to highlight the extremes of gust frequencies - low feedback of $f=5$ Hz in environment 2, high feedback of $f=51$ Hz in environment 3, and no feedback in regime 1. All three share the $26$ Hz frequency signature expressed by wing oscillation. It is expected that some environments will be identified with higher accuracy than the others, and strain dynamics from each environment are sparsely classified separately to reveal this structure in the data.

For each flow environment $i$, the tests are conducted using tenfold cross validation. Ten percent of vectors in $\hat\bS_i$ are randomly selected to be test inputs of strain dynamics, and the remaining 90\% of the columns are used to train the POD library and optimal sensors, \revision{meaning the POD modes from each environment are trained from 500 observations.}{} The POD library $\hat\bPsi$ consists of the stacked dominant POD modes of environments 1,2 and 3 shown in Figure \ref{fig:fouriermodes}. After training the POD library, test frequency dynamics $\hat\bs\in\reals^{256}$ are classified using $L_1$ constrained approximation of the overdetermined linear system\eqref{eqn:src_objective2}. The performance is investigated across increasing levels of sensor noise, and the classification accuracies across all the tests for one level of sensor noise are displayed as one box and whisker in the subsequent plots. The mean, 25th quartile, and 75th quartile of the classification performance distribution are displayed as circles, the bottom, and top edges of the boxes, respectively. Whiskers extend to $\pm 2.7$ standard deviations of the result distribution, and any data outside this range is displayed as a small outlier point. 

Box and whisker plots of classification accuracy for test dynamics from environments 1-3 are shown in Figure \ref{fig:reg123_l1}. $L_1$ constrained approximation achieves perfect 100\% classification accuracy in the noiseless case for all three environments, and even at levels of 20\% sensor noise, the performance remains impressively higher than chance \revision[R2:22]{ or 33.3\%, the scenario in which the classifier chooses between the three environments uniformly at random.}{(33\%)} Interestingly, as sensor noise increases, the identification of high-frequencies in environment 3 is more successful than the identification of environments 1 and 2. Environment 3 in particular contains high frequency content and higher amplitude spatial modes (Figure \ref{fig:strainplots}) that are amplified in the frequency domain analysis.

These results are quite sensitive to the choice of $L_2$ error tolerance $\epsilon$. In these computations, $\epsilon$ is set to some multiple of the least squares solution error. That is, given the cheaply computed least squares solution $\ba_{ls}=\argmin_{\tilde\ba}\|\hat\bPsi\tilde\ba-\hat\bs\|_2$, $\epsilon_{ls}=\min_{\tilde\ba}\|\hat\bPsi\tilde\ba-\hat\bs\|_2$. In practice, setting $\epsilon=\epsilon_{ls}$ in the noiseless case and relaxing the tolerance to $\epsilon=1.2\epsilon_{ls}$ in the noisy case works quite well. The tolerance appears to depend on the noise level $\eta$ and $\epsilon_{ls}$, and there may exist an optimal choice based on the two parameters requiring further investigation.

\begin{figure}
\centering
\begin{overpic}[width=.9\textwidth]
{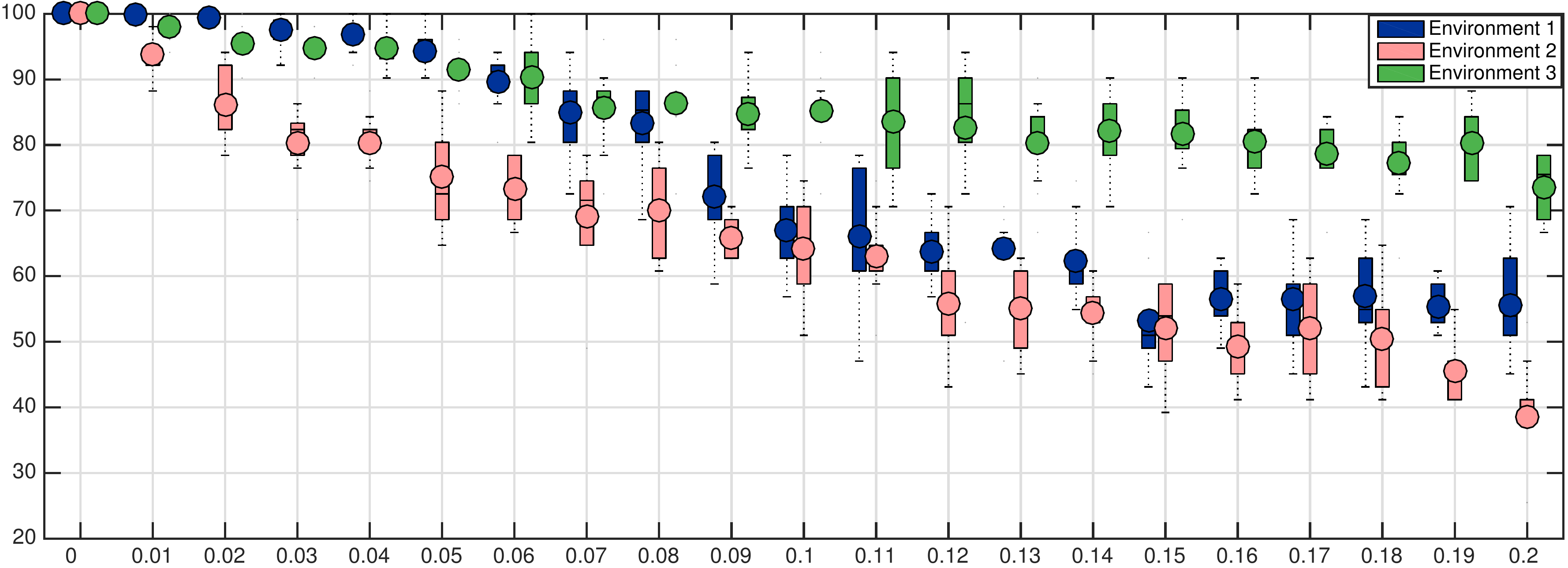}
\put(-2,20){\makebox(0,0){\rotatebox{90}{\% Accuracy}}}
\put(40,-2.5){\small Sensor noise variance $\eta$}
\end{overpic}
\vskip 2mm
\caption{The cross-validated performance of $L_1$ approximation across environments 1-3 is shown here for sensor noise of increasing variance $\eta$. Red, blue and green represent input observations from environments 1,2 and 3, respectively. High-frequency gusts from regime 3 are more robustly distinguished from the other two regimes.\label{fig:reg123_l1}}
\end{figure}
\begin{figure}
\centering
\begin{overpic}[width=.32\textwidth]
{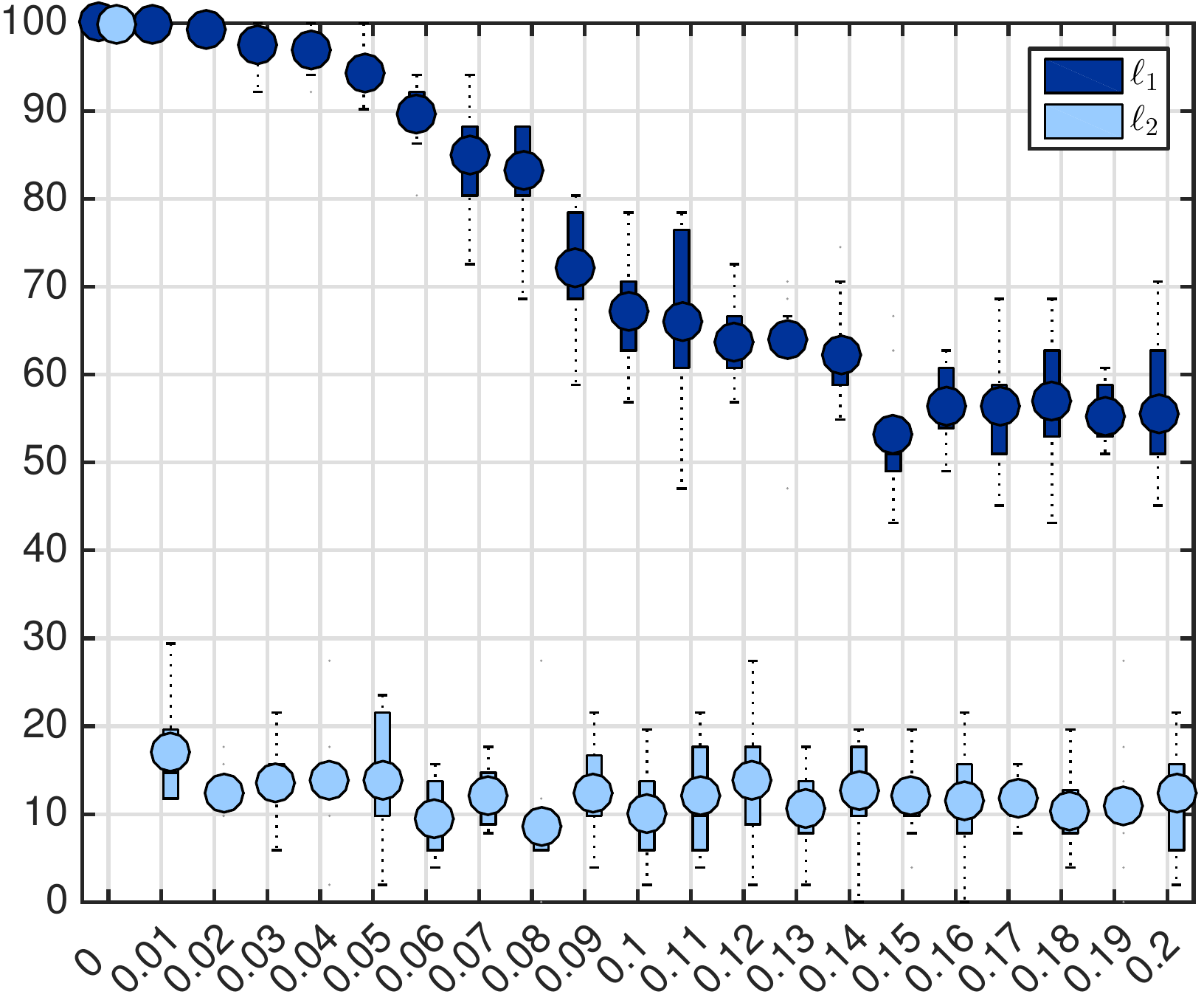}
\put(40,75){\small Environment 1}
\put(-5,40){\makebox(0,0){\rotatebox{90}{\% Accuracy}}}
\end{overpic}
\begin{overpic}[width=.32\textwidth]
{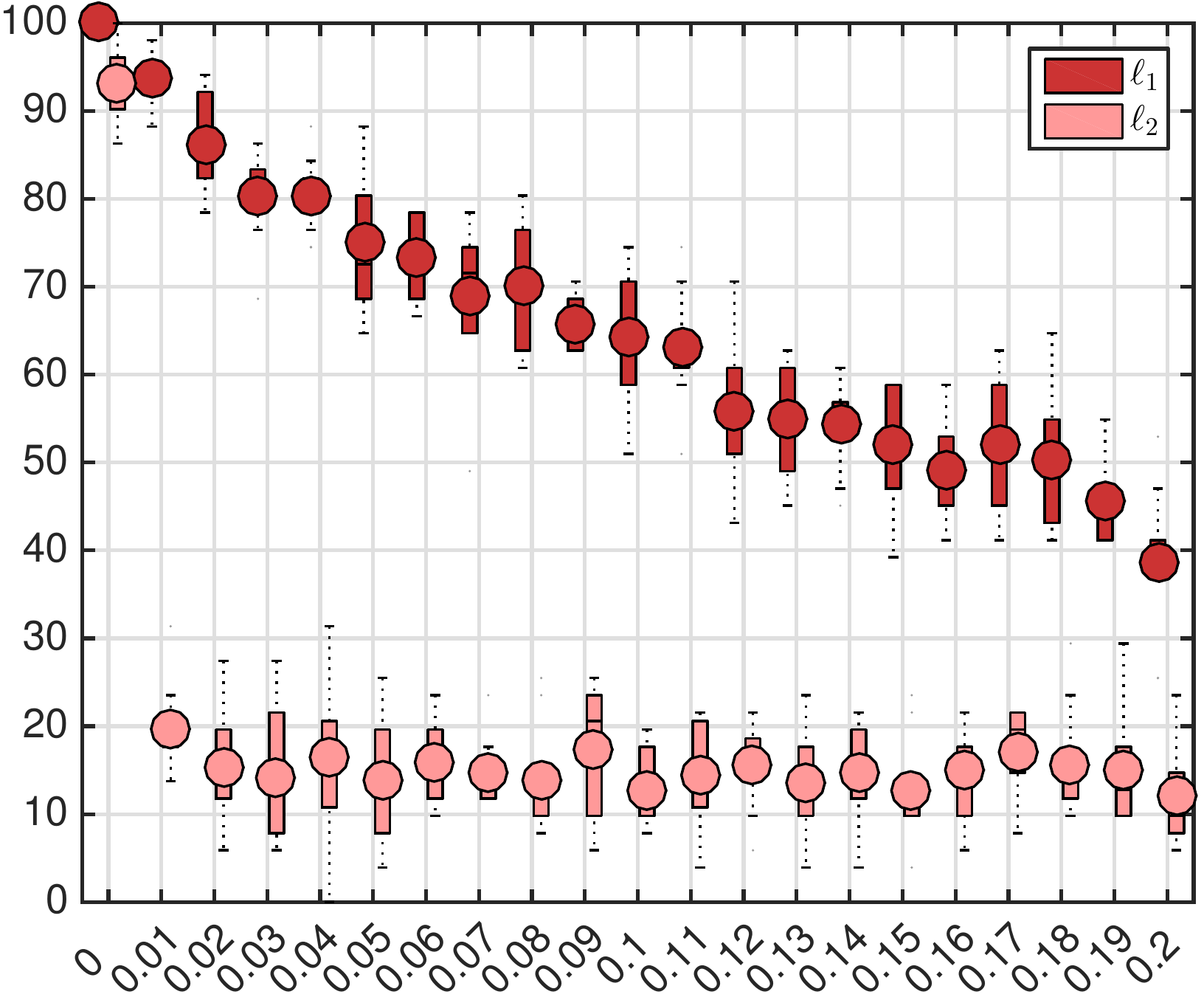}
\put(40,75){\small Environment 2}
\put(20,-7){\small Sensor Noise Variance $\eta$}
\end{overpic}
\begin{overpic}[width=.32\textwidth]
{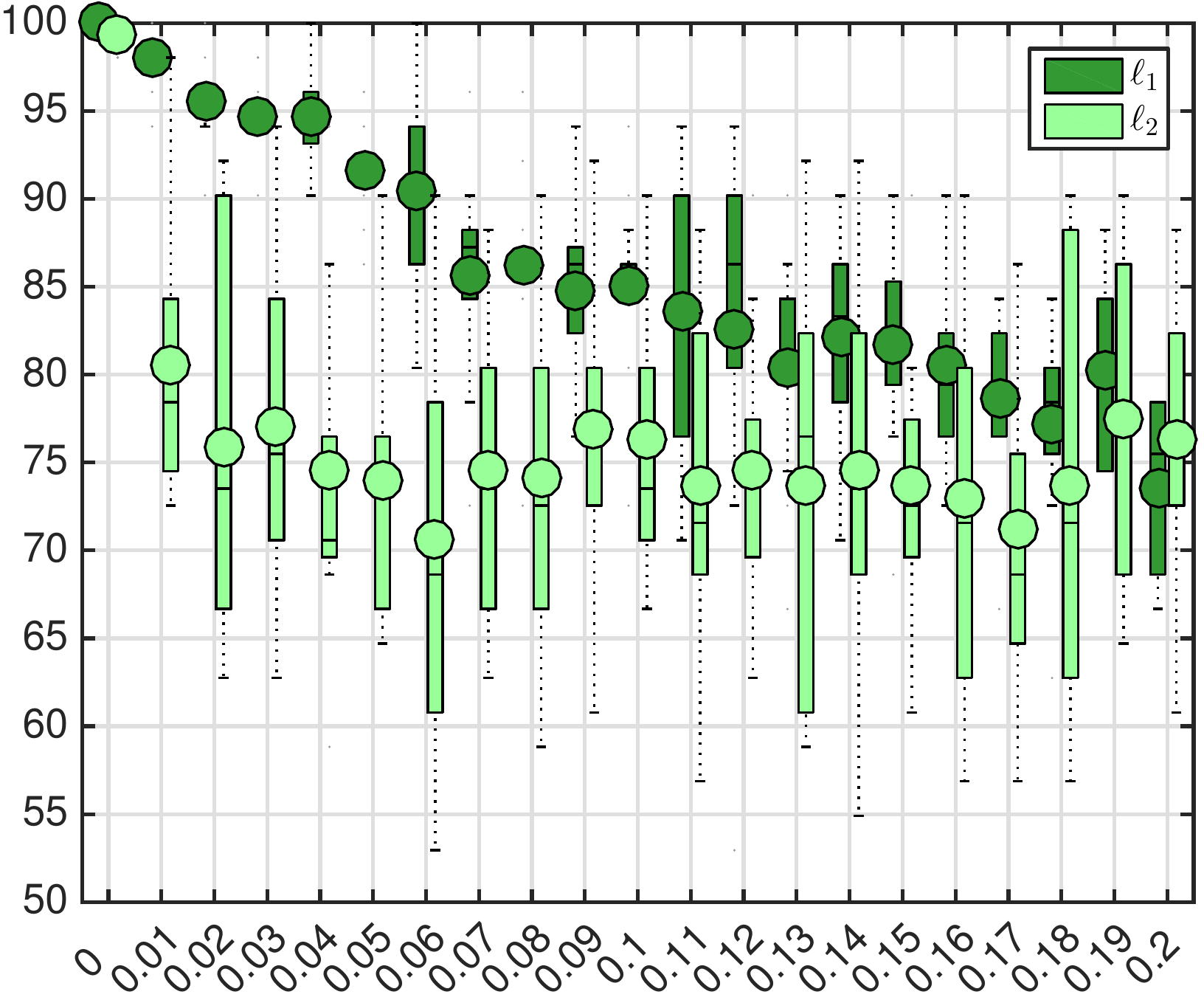}
\put(40,75){\small Environment 3}
\end{overpic}
\vskip 2mm
\caption{The comparison of $L_1$ approximation to least squares approximation of the library coefficients demonstrates the power of the sparsity promoting $L_1$ norm constraint for subset selection. \label{fig:SRC_least_squares}}
\end{figure}

\subsection{Sparse ($L_1$) vs. least squares ($L_2$) approximation } \label{ssec:srcf_leastsq}
It is instructive to overlay in Figure \ref{fig:SRC_least_squares} results from classification using the least squares solution $\ba_{ls}$ in \eqref{eqn:classifier}. The comparison demonstrates the clear advantage of using $L_1$ norm minimization of library coefficients over $L_2$ minimization of the residual (least squares), particularly in the presence of sensor noise, although least-squares achieves $L_1$ accuracy in the noiseless case. The identification of environment 3 is once again an exception due to high-frequency fitting that is not present in the other two, however, $L_1$ constrained approximation always outperforms least squares. Therefore the sparsity promotion of solution library coefficients is essential for effective classification in the discriminating frequency domain.

\begin{figure}
\centering
\begin{overpic}[width=.45\textwidth]
{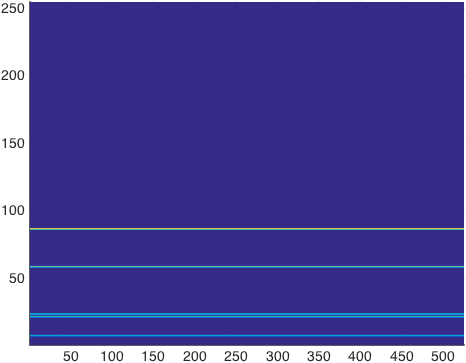}
\put(-5,40){\makebox(0,0){\rotatebox{90}{Selected frequencies}}}
\put(40,73){\color{white}5 QR selections}
\end{overpic}
\begin{overpic}[width=.45\textwidth]
{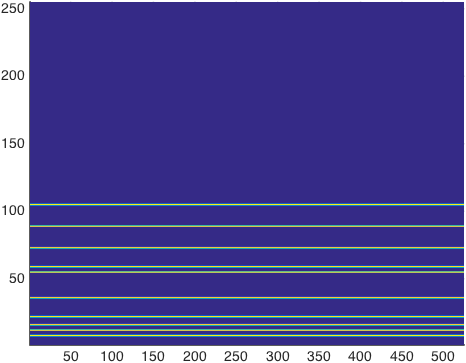}
\put(40,73){\color{white}10 QR selections}
\end{overpic}
\vskip 2mm
\caption{Ensemble of all QR frequency selections from nearly five-hundred instances of cross-validation, that is, QR selection using POD modes trained from five-hundred random observations. Note that optimal sampling locations are stable and experience no variation across different training sets. \label{fig:qdeimsensors}}
\end{figure}

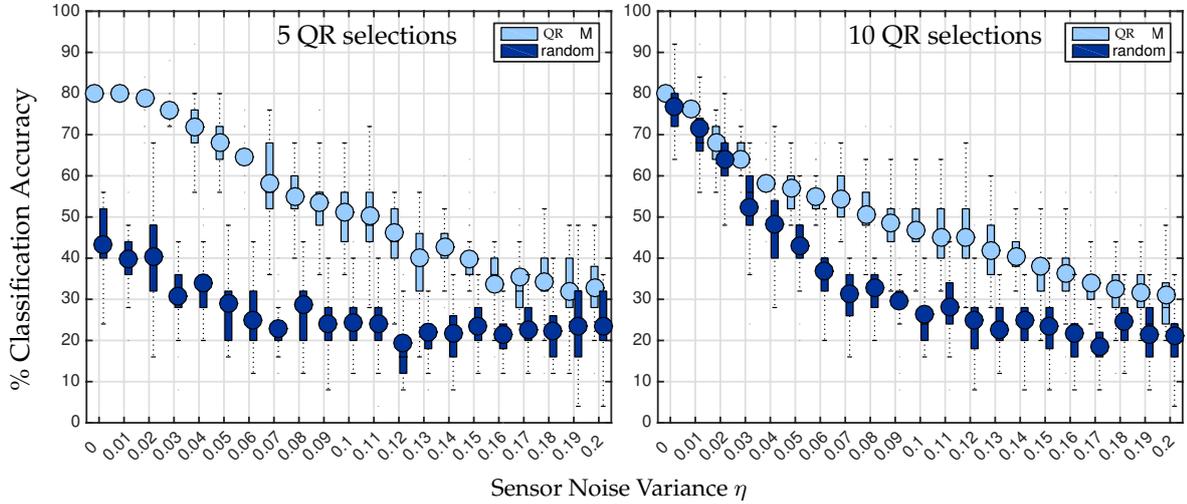
\begin{figure}
\centering
\setlength{\fboxrule}{0pt}%
\begin{overpic}[width=.45\textwidth]
{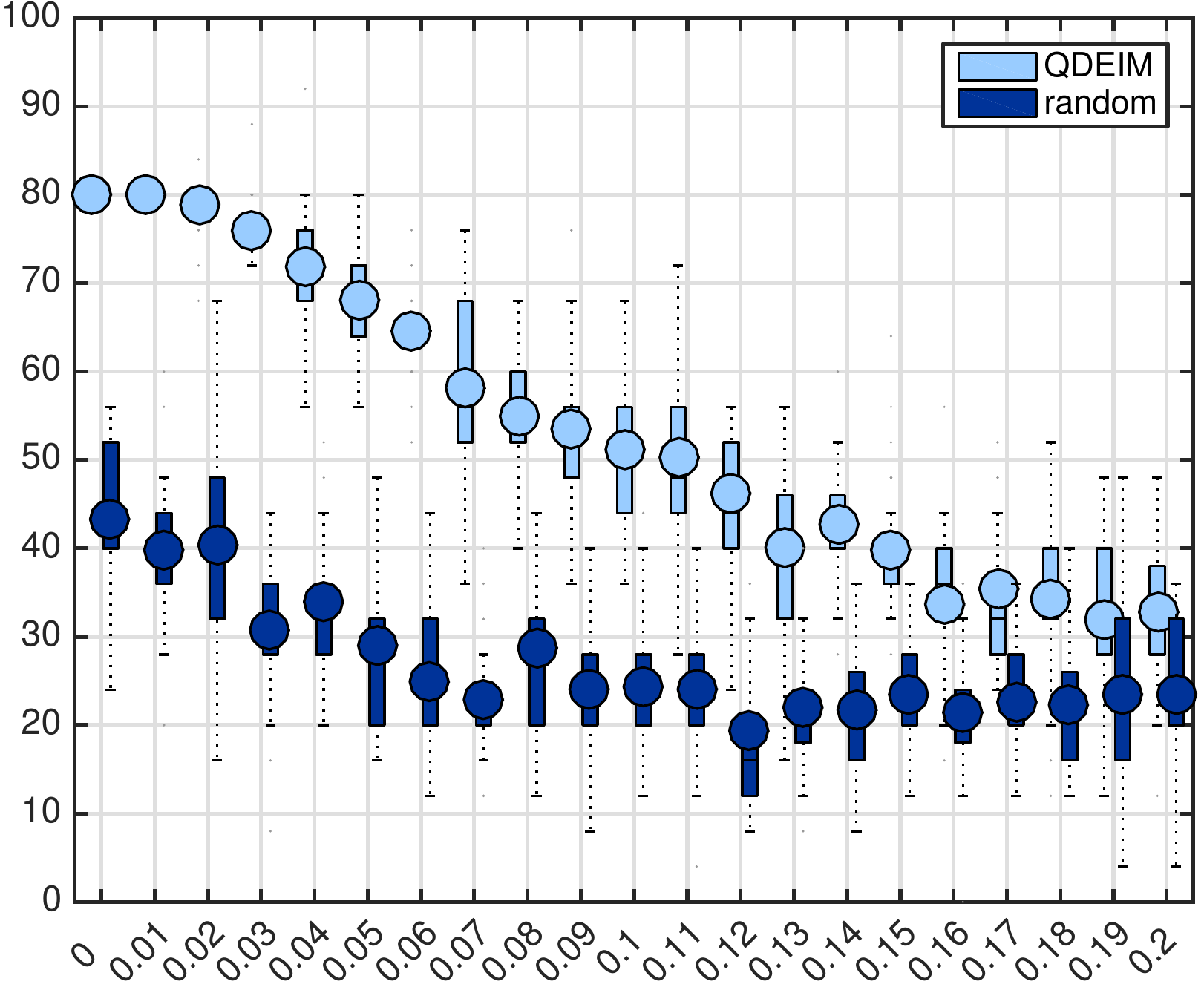}
\put(-5,40){\makebox(0,0){\rotatebox{90}{\% Classification Accuracy}}}
\put(86.5,76){\fcolorbox{white}{white}{\tiny QR\quad}}
\put(40,75){5 QR selections}
\end{overpic}
\begin{overpic}[width=.45\textwidth]
{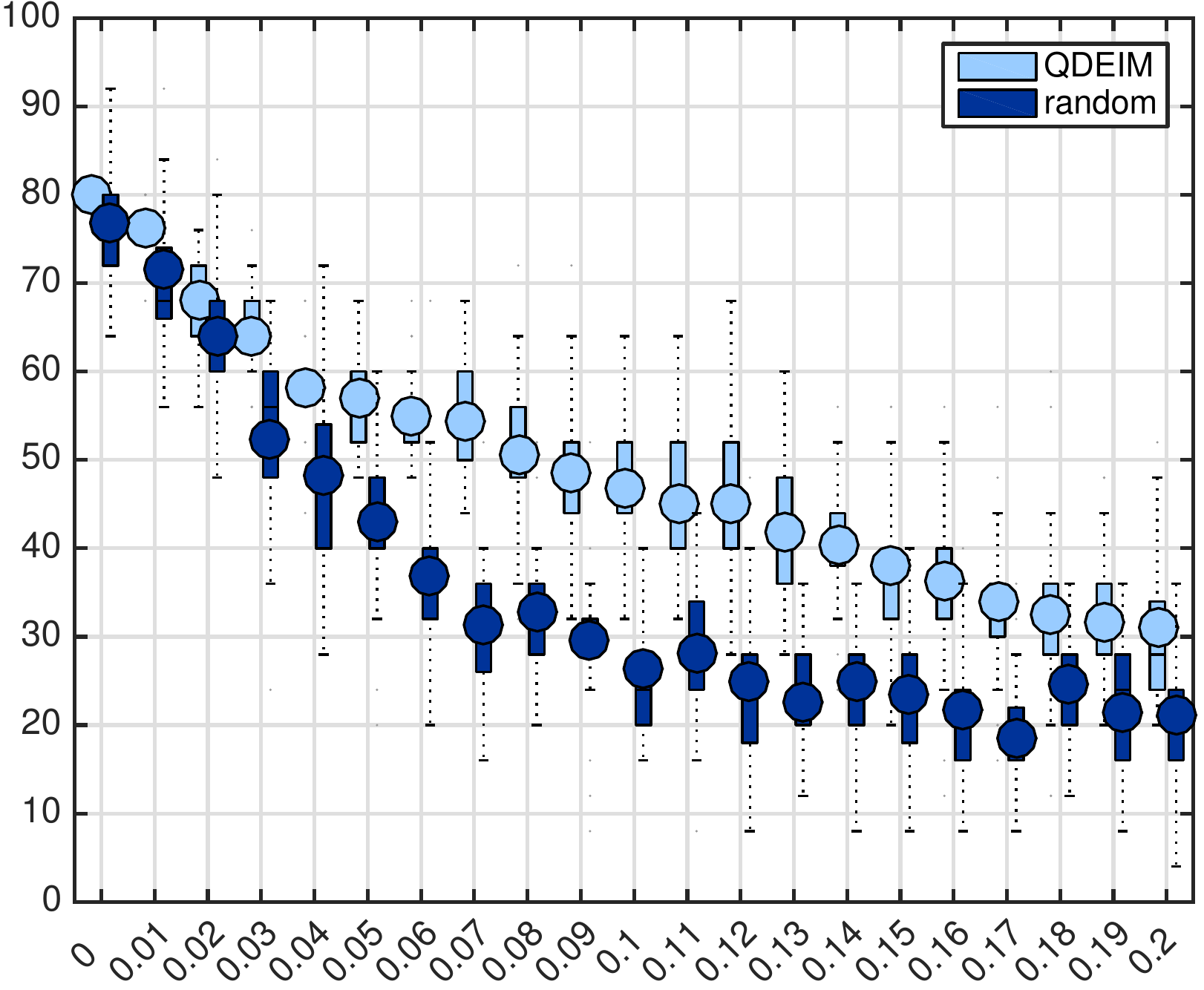}
\put(86.5,76){\fcolorbox{white}{white}{\tiny QR\quad}}
\put(40,75){10 QR selections}
\end{overpic}

{\small  Sensor Noise Variance $\eta$}
\caption{{\bf SRCf Accuracy}: This comparison of \revision{QR selected frequencies}{Q-DEIM frequency samples} with random frequency samples demonstrates that even a 2\% (5 row) \revision{QR selection}{Q-DEIM samples} outperforms a 14\% random row selection (35 rows) for classifying five environments for increasing sensor noise. \label{fig:qdeimrandomcomp}}
\end{figure}

\subsection{SRCf with \revision{QR selection}{Q-DEIM sampling}} \label{ssec:csrcf_qdeim}

The resolution of strain measurements in time is directly informed by the optimized sampling of frequency content for classification. The strain dynamics from different flow environments by construction are linearly separable by flow feedback frequency. The precise discriminating frequencies are unknown in the training stage but are easily determined from \revision{QR}{the Q-DEIM} selection with the joint POD of all strain dynamics. 
The following simulation compares the performance of SRCf with \revision{QR row selection}{Q-DEIM samples} over random samples in the frequency domain, confirming the advantage of \revision{QR selection}{Q-DEIM selection samples}. 

\revision{QR selection is}{Q-DEIM samples are} obtained from a richer training dataset of strain dynamics consisting of forced disturbances at 25 Hz, 50 Hz, 75 Hz and 100 Hz, in addition to the no forcing scenario. There are now five environments considered by the classifier, a more difficult task than the former. As before, the tested dynamics do not consist of the same random 90\% of the data used for training POD library modes and \revision{QR selection}{Q-DEIM samples}. The five \revision{QR selected measurements}{Q-DEIM samples} for the five environment case in Figure \ref{fig:qdeimsensors} appear to cluster around the discriminating disturbance frequencies and even appear to discriminate the flow forcing at 25 Hz from wing oscillation at 26 Hz. The measurements do not exactly lie at these numbers because the DCT does not span whole number frequencies. 

The performance of SRCf shown in Figure \ref{fig:qdeimrandomcomp} demonstrates the advantage of \revision{QR selection}{Q-DEIM samples} of strain dynamics over random sampling. Up to a 35\% increase in classification accuracy is achieved with only five samples (1.95\% of the 256 element vector of strain dynamics). In addition, \revision{QR selection is}{Q-DEIM samples are} more robust to sensor noise than an even larger number of random samples, although the accuracy gain drops off near sensor noise of level $\eta=0.2$ that comprises 20\% of the untransformed signal's magnitude. The strain dynamics are saturated by noise at this higher noise level where SRCf performance drops off to randomized classification accuracy.

Interestingly, doubling the number of \revision{QR selections}{Q-DEIM samples} does not improve the accuracy, although doubling the number of random samples improves random performance as expected. \revision[R1:7]{The results indicate that cross-environment variation is well characterized by the QR selected measurements which are optimized for library approximation with $\hat{\bPsi}$.}{Therefore, the five \revision{QR selections}{Q-DEIM samples} optimally encode the discriminating feedback frequencies necessary for classification.} Furthermore, increasing the number of random samples to 35 (Figure \ref{fig:resultschordplacement}) does not achieve the same robust accuracy drop-off seen with \revision{QR selections}{Q-DEIM samples}. 

\begin{figure}
\centering
\begin{overpic}[width=.95\textwidth]
{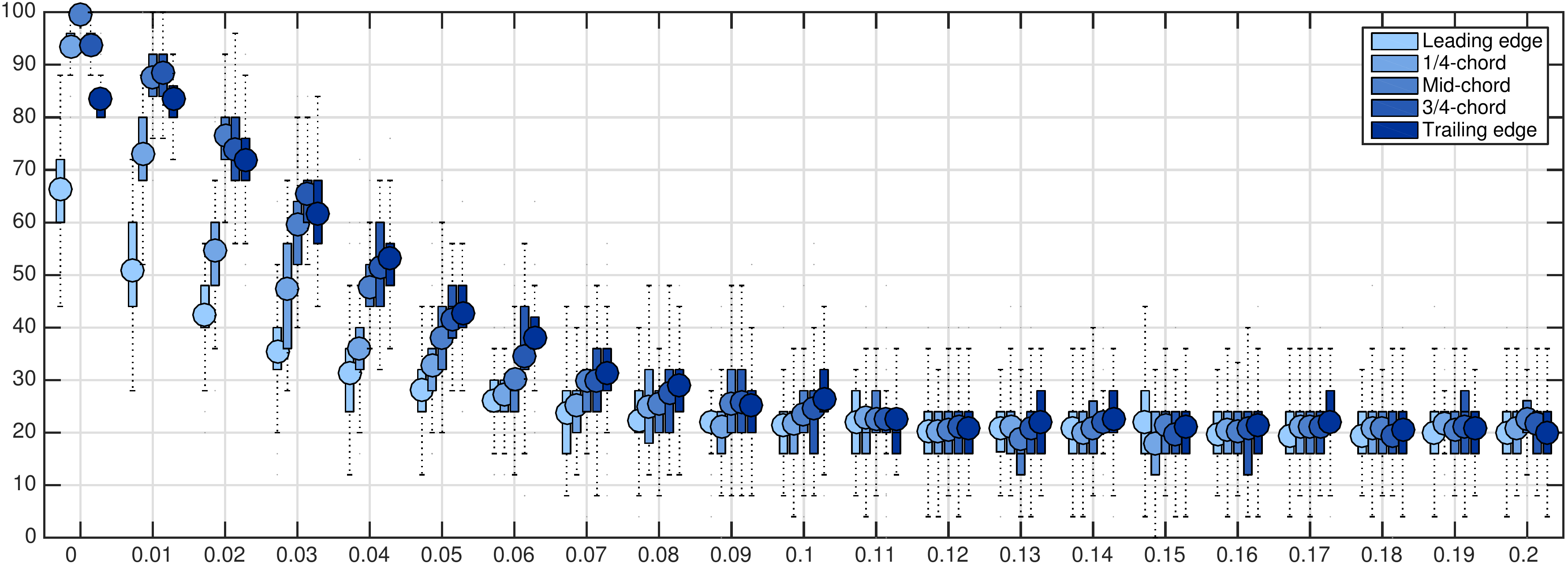}
\put(-2,20){\makebox(0,0){\rotatebox{90}{\% Accuracy}}}
\end{overpic}

{\small Sensor noise variance $\eta$}
\caption{This plot compares classification accuracy for sensors distributed around (left to right) leading edge, $1/4$ chord, mid-chord, $3/4$-chord and trailing edge for identically increasing sensor noise. Each error bar represents cross-validated classification accuracy across observations randomly chosen from a Gaussian distribution centered around a certain area of the chord, sampled from a randomly chosen environmental regime out of the five total regimes. Classification accuracy increases across the chord with accuracy increasing near the trailing edge.\label{fig:resultschordplacement}}
\end{figure}

\subsection{SRCf with spatial bias} \label{ssec:csrcf_chordwise}

A related inquiry concerns the optimal spatial locations along the wing to capture strain dynamics. These locations would correspond to the physical placement of each strain sensor and campaniform sensilla on insect wings.
Our previous simulations ignored any spatial correlation by testing classification for frequency content from random locations along the chord (columns of $\hat\bS$). We now introduce a spatial bias that draws strain frequency dynamics at random from Gaussian distributions centered at different regions of the wing chord and training with the remainder, using the same data considered previously in \cref{ssec:csrcf_qdeim}. This essentially tests columns $\hat\bs$ that are adjacent in the data matrix $\hat\bS$. The different regions of the chord considered - the leading edge, 1/4-chord, midchord, 3/4-chord, and trailing - are important in the study of both insect wings and airfoils. The SRCf classifier is used with 35 random frequency samples in the frequency domain. 

The expected performance profile for increasing sensor noise is observed in the results of Figure \ref{fig:resultschordplacement}, however there is an obvious classification advantage towards the trailing edge of the wing. This trend is consistent with downstream amplification of flow-induced strain towards the trailing edge, where the wing is most prone to deflection from the free boundary \eqref{eqn:beamboundary} and flexural stiffness is lowest. However, this finding deviates from observed campaniform sensilla locations that tend to be distributed away from the trailing edge. There are many possible explanations - the trailing edge is dominated by membrane surface without venal or neural conduits for the sensilla, or the non-stiff trailing edge of the wing may be prone to amplified flow process noise. Alternatively, the ensemble of campaniform sensilla may be sufficient to aggregate strain dynamics to amplify flow features without resorting to trailing edge sensors. Both scenarios are well worth examining in future research.

\section{Conclusions and Outlook}

This work develops a data-driven framework for flow environment identification based on supervised learning from simulated wing strain dynamics in the frequency domain. This is done using sparse classification in an overcomplete library of POD modes, which is assembled by simulating strain dynamics from each expected flow environment. Then, the problem of identifying the flow environment reduces to a classification task between the sets of POD modes in the library. The subset selection is accomplished by $L_1$ regularized, sparse approximation of POD library coefficients from input frequency dynamics. This $L_1$ regularization is extremely accurate and mitigates contributions from noise when compared to least squares approximation -- even with sensor noise at 20\% of the input signal's magnitude, sparse classification is slightly biased towards the correct originating environment. Furthermore, SRCf with strategically subsampled signals is shown to be effective, with discriminating frequency measurements selected by the pivoted QR factorization of POD modes.
In this manner, sparse classification of strain frequencies facilitates environment identification from single wing locations, providing insight into possible sparse encoding strategies employed by strain sensing neurons on insect wings.
Moreover, our approach may be applied to sensor-equipped feedback systems in flight control, which is an active area of research. 

\subsection{Control Implications}
Downstream environment identification from sensor data is an important problem in closed-loop control where sensors are expensive and measurements are corrupted by noise. Spatial strain mode shapes do not vary significantly across environments and require many more physical strain sensors to complete spatial knowledge. Frequency domain analysis of wing strain can help move toward autonomous flight technologies. Furthermore, the design of frequency samples chosen specifically to discriminate certain environments or frequencies greatly reduces the dimension of decision space and helps mitigate uncertainty in sensor measurements. \revision[R2:25]{Strain sensors and gauges are already being used to characterize stresses and guide wing and fin design in experiment \citep{KahnTangorra:2015}. As an added functionality they may assist in flow characterization using similar data-driven dimension reduction concepts as presented here.}{Indeed, strain sensors or gauges are commonly used in miniaturized aircraft for readily available structural encodings of surrounding flows.} Our frequency distinction framework is particularly attractive for feedback analysis for controller decisions. \revtwo[R2:4]{Given suitable libraries of candidate frequency dynamics, the sparse classification framework can be efficiently implemented in flight controllers, especially when the decision space consists of only a few strategically selected frequencies. For example, controllers may mitigate undesirable frequencies encountered in flight such as the erratic high frequency content of turbulence and low frequencies induced by strong winds. Although we have not built a controller in this work, we note that convex optimization strategies, such as the one underlying sparse classification, are commonly employed in control applications.} 

\subsection{Biological Implications}
There are a number of intriguing biological conjectures that the current results help address.
Specifically, the machine learning architecture of learned dictionaries and sparse sampling both
seem to have advantageous properties in the context of flight.  The learned libraries encoded from
our numerical simulations suggest that insects possess a similar, innate library for flight dynamics.
Indeed, insects at birth simply begin flying without a lengthy training stage, suggesting that
flight dynamics, or dictionaries, are \revision[R2:23]{stereotyped behavior}{hard-wired} in the mechanosensory flight system \revision[R2:23]{\citep{cole1982pattern,gettrup1966sensory,Dickerson:2014}}{}.  Thus at birth, a low-dimensional representation of mechanosensory codes are genetically inherited.  The sparse and stereotyped placement of campaniform sensilla on the wings of a hawkmoth also suggest that sparse sampling does indeed occur for helping guide flight dynamics and control protocols.  Of course, we have made no connection to control, but we certainly have demonstrated the tremendous bio-inspired advantages of sparsity and learned libraries.

\subsection{Outlook}
Here we propose several avenues of investigation aimed towards engineering and biology applications that exploit low-rank data structure. One direction concerns detecting and adapting this supervised learning and sparse classification framework to new, unseen classes of strain dynamics. In particular, turbulent and mixed frequency flow environments may introduce nonlinear decision boundaries between classes that may require a boosted classifier that could, for example, aggregate decisions from multiple spatial locations on the wing. The supervised library learning stage may be generalized to assemble modal representations other than POD, such as dynamic mode decomposition (DMD). The second component of measurement selection, empirical interpolation methods, would benefit from incorporation of measurement constraints that typically arise in engineering applications. 
	
Furthermore, both components of this framework, sparse classification and measurement selection, stand to benefit from the quantification of uncertainty in environment identification. The assumption of white Gaussian sensor noise in our work is quite simplistic; more realistic sources of uncertainty may arise from process noise such as uncertainty in training data or spatially varying amplification of noise. Since this is an equation-free, data-driven approach to environment identification, more sophisticated learning strategies such as neural networks, decision trees, etc. may be employed in the training stage to identify and mitigate these sources of uncertainty.}


\revision[R1:9]{This work is a piece that fits into a broader effort to understand neurosensory encoding for robust insect flight.  
In the present effort, a coupled fluid-structure interaction model was used to simulate instead loads on an insect-scale wing.  
However, applying this analysis to measurements of the strain on an actual wing presents an exciting avenue of future research.  
With increasingly small and inexpensive strain sensors, it may be possible to explore optimal sensor location in a robotic platform.  
Perhaps of greater interest, it may be possible in future studies to incorporate data directly from individual campaniform sensilla in a flying insect.  
It would also be interesting to compare the distribution of campaniform sensilla across insect wings to attempt to understand underlying biological optimization principles and goals.  
}{}

\section*{Acknowledgements}
We are grateful to Bingni Brunton, Josh Proctor and Ido Bright for invaluable conversations
relating to sparsity and library learning. We also thank Tom Daniel, Annika Eberle, Brad Dickerson and Thomas Mohren for fruitful discussions on strain sensing on insect wings. 
 K. Manohar acknowledges the support of a Seattle ARCS Foundation fellowship. 
 S. L. Brunton acknowledges support from the Air Force Research Labs (FA8651-16-1-0003) and the Air Force Center of Excellence on Nature Inspired Flight Technologies and Ideas (FA9550-14-1-0398).  
 J. N. Kutz acknowledges support from the Air Force Office of Scientific Research
 (FA9550-15-1-0385).  
 SLB and JNK also acknowledge support from the Defense Advanced Research Projects Agency (DARPA contract HR0011-16-C-0016).  

\setcounter{section}{0}
\renewcommand\thesection{\Alph{section}}

\begin{spacing}{.89}
\footnotesize{
\setlength{\bibsep}{3pt}
\bibliographystyle{elsarticle-harv}
\bibliography{MAIN_wing_final}}
\end{spacing}
\end{document}